\pdfoutput=1
\documentclass[usenatbib]{mn2e}
\usepackage[implicit=false,breaklinks,colorlinks,citecolor=blue]{hyperref}
\usepackage{apjfonts}
\usepackage{amssymb}
\usepackage{amsmath}
\usepackage{ctable}
\usepackage{url}
\usepackage{comment}
\usepackage{breakurl}
\usepackage{fixltx2e} 

\newcommand{\be}{\begin{equation}}
\newcommand{\ee}{\end{equation}}

\newcommand{\paperone}{Paper {\small I}}
\newcommand{\papertwo}{Paper {\small II}}

\newcommand{\gizmourl}{\burl{http://www.tapir.caltech.edu/~phopkins/Site/GIZMO.html}}
\newcommand{\vspacerpostplot}{\vspace{-0.4cm}}

\newcommand\plotonesize[2]
 {\centering \leavevmode \includegraphics[width={#2\columnwidth}]{#1}}
\newcommand{\plotsidesize}[2]
 {\centering \leavevmode \includegraphics[width={#2\textwidth}]{#1}}
\newcommand{\acknowledgments}{\begin{small}\section*{Acknowledgments}\end{small}}
\newcommand\altaffilmark[1]{$^{#1}$}
\newcommand\altaffiltext[1]{$^{#1}$}
\title[Mesh-Free Anisotropic Diffusion]{Anisotropic Diffusion in Mesh-Free Numerical Magnetohydrodynamics\vspace{-0.5cm}}

\author[Hopkins]{
\parbox[t]{\textwidth}{ 
Philip F. Hopkins\altaffilmark{1}\thanks{E-mail:phopkins@caltech.edu}
} 
\vspace*{6pt} \\
\altaffiltext{1}{TAPIR \&\ The Walter Burke Institute for Theoretical Physics, Mailcode 350-17, California Institute of Technology, Pasadena, CA 91125, USA\vspace{-1.1cm}} \\
}

\date{Submitted to MNRAS, January, 2016\vspace{-0.6cm}}
\begin{document}
\maketitle
\label{firstpage}

\begin{abstract}
We extend recently-developed mesh-free Lagrangian methods for numerical magnetohydrodynamics (MHD) to arbitrary anisotropic diffusion equations, including: passive scalar diffusion, Spitzer-Braginskii conduction and viscosity, cosmic ray diffusion/streaming, anisotropic radiation transport, non-ideal MHD (Ohmic resistivity, ambipolar diffusion, the Hall effect), and turbulent ``eddy diffusion.'' We study these as implemented in the code {\small GIZMO} for both new meshless finite-volume Godunov schemes (MFM/MFV). We show the MFM/MFV methods are accurate and stable even with noisy fields and irregular particle arrangements, and recover the correct behavior even in arbitrarily anisotropic cases. They are competitive with state-of-the-art AMR/moving-mesh methods, and can correctly treat anisotropic diffusion-driven instabilities (e.g.\ the MTI and HBI, Hall MRI). We also develop a new scheme for stabilizing anisotropic tensor-valued fluxes with high-order gradient estimators and non-linear flux limiters, which is trivially generalized to AMR/moving-mesh codes. We also present applications of some of these improvements for SPH, in the form of a new integral-Godunov SPH formulation that adopts a moving-least squares gradient estimator and introduces a flux-limited Riemann problem between particles. 
\end{abstract}

\begin{keywords}
methods: numerical --- hydrodynamics --- MHD --- instabilities --- conduction --- diffusion
\vspace{-1.0cm}
\end{keywords}

\vspace{-1.1cm}
\section{Introduction}
\label{sec:intro}

Diffusion operators are ubiquitous in physical systems, manifest in e.g.\ conduction, viscosity, cosmic ray diffusion/streaming, photon diffusion (in optically thick media), sub-grid ``turbulent eddy'' diffusion, passive scalar (e.g.\ metal) diffusion, Ohmic resistivity, ambipolar diffusion, the Hall effect, and more. Most of these are fundamentally anisotropic. Magnetic fields, fluid flow, or radiation flux can all break isotropy and introduce a (local) preferred direction. The effects are large -- for the same initial conditions but different magnetic field configurations, effects like viscosity or conduction might dominate the hydrodynamics, or be completely suppressed. Moreover, anisotropy can introduce qualitatively new behaviors and physical instabilities \citep[e.g.][]{balbus.2000:mti,quataert.2008:hbi}. 

Clearly, it is important to capture these phenomena accurately in numerical simulations. In regular, non-moving mesh methods (including fixed-grid and adaptive mesh-refinement [AMR]), the properties of diffusion equations are reasonably well-understood, but even in these cases it is highly non-trivial to formulate anisotropic operators in a numerically stable fashion \citep[see, e.g.][]{sharma.2007:anisotropic.diffusion.stability.limiters}. But for some classes of problems, these methods are sub-optimal. They tend to produce excessive diffusion when a fluid is moving rapidly relative to the grid, especially across contact discontinuities \citep{tasker:2008.gas.turb.vs.gal.prop,springel:arepo,hopkins:gizmo}; have difficulty coupling to N-body gravity methods and handling self-gravitating hydrostatic equilibrium \citep{muller:1995.grid.code.gravity.problems,leveque:1998.godunov.source.term.balance,zingale:2002.grid.hydro.eqm.issues}; introduce low-order errors around refinement boundaries \citep{oshea:sph.tests,heitmann:2008.cosmic.code.comparison}; and feature inherently preferred directions which can introduce systematic errors even at high resolution if physical anisotropies are not aligned with the grid axes \citep[e.g.][]{peery:carbuncle.discovery,hahn:2010.disk.gal.orientations.ramses,hopkins:gizmo,hopkins:mhd.gizmo}. 

Mesh-free methods can avoid these sources of error, and so are popular for many problems in astrophysics. However, the most popular mesh-free method, smoothed-particle hydrodynamics (SPH) has its own difficulties. For example, SPH requires the use of artificial diffusion operators to maintain numerical stability, which can produce greater numerical viscosity and over-smoothing of shocks and discontinuities compared to Riemann-based shock solvers \citep{cullen:2010.inviscid.sph}; it tends to suppress fluid-mixing instabilities \citep{morris:1996.sph.stability,ritchie.thomas:2001.egy.wtd.sph,agertz:2007.sph.grid.mixing}; and suffers from zeroth-order errors \citep[``E0'' errors;][]{morris:1996.sph.stability,dilts:1999.sph.stability,read:2010.sph.mixing.optimization} which produce systematic errors that do not converge with resolution alone. Although there have been dramatic improvements in all of these \citep[see][]{hopkins:lagrangian.pressure.sph,rosswog:2014.sph.accuracy,hu:2014.psph.galaxy.tests}, the E0 errors cannot be completely eliminated from SPH without making the method numerically unstable \citep{price:2012.sph.review}. 

Recently, however, \citet{lanson.vila:2008.meshfree.consistency,lanson.vila:2008.meshfree.convergence,gaburov:2011.meshless.dg.particle.method,hopkins:gizmo} have developed a class of new, mesh-free Lagrangian finite-volume methods which are both high-order consistent and fully conservative. Similar to moving-mesh codes \citep{springel:arepo,duffell:2011.TESS,gaburov:2012.public.moving.mesh.code}, these new methods appear to capture many of the advantages of AMR and SPH, while avoiding the disadvantages above (although, of course, they feature their own sources of error such as enhanced ``grid noise'' and volume ``partition noise''; see \citealt{hopkins:gizmo}).  In \citealt{hopkins:gizmo} (hereafter \paperone) and \citet{hopkins:mhd.gizmo} (\papertwo), these are developed for magnetohydrodynamics (MHD) in the multi-method magnetohydrodynamics+gravity code {\small GIZMO},\footnote{A public version of the code is available at \gizmourl}  built on the $N$-body gravity and domain decomposition algorithms from {\small GADGET-3} \citep{springel:gadget}.

In this paper, therefore, we extend these new methods to include arbitrary anisotropic diffusion operators. We consider a systematic comparison of tests in order to determine the degree of anisotropy that can be reliably treated. In the Appendices, we show how some parts of the new methods here can also be applied to SPH.

\vspace{-0.5cm}
\section{General Diffusion Operators in Meshless Finite-Volume Godunov Methods}
\label{sec:mfm}

The general diffusion equation in conservative form is:
 \begin{align}
\frac{d{\bf U}}{d t} &=  -\nabla \cdot {\bf F}_{\rm diff} \\
{\bf F}_{\rm diff} &= -{\bf K} \cdot \left( \nabla \otimes {\bf q}  \right)
 \end{align}
where $d/dt$ is a Lagrangian derivative, and ${\bf F}_{\rm diff}$ a diffusive flux given by a linear combination of the tensor field ${\bf K}$ and the gradients of the field ${\bf q}$. We denote the inner product ``$\cdot$'' and outer product ``$\otimes$.'' 

In the mesh-free finite-volume or finite-mass (MFV or MFM, respectively) formulations of the Lagrangian methods in \citet{gaburov:2011.meshless.dg.particle.method} and \paperone, we begin from a general conservation equation of the form $D{\bf U}/Dt = -\nabla \cdot {\bf F}$ and use this to derive a second-order accurate Godunov-type numerical expression of the hydrodynamic equations. It is therefore trivial to apply the same here, giving
\begin{align}
\frac{d}{dt}(V\,{\bf U})_{i} &= - \sum_{j}\,{{\bf F}}_{{\rm diff},\,ij}^{\ast}\cdot {\bf A}_{ij} \label{eqn:faces}
\end{align}
where ${\bf A}_{ij}$ is the ``effective face area'' (defined in \paperone; it depends on the inter-element spacing and kernel shape, and reduces to the geometric faces of a Voronoi tesselation in the limit of a delta-function kernel). The same would obtain for Cartesian grid methods and moving-mesh codes, with ${\bf A}_{ij}$ the usual inter-face area. Here ${{\bf F}}_{{\rm diff},\,ij}^{\ast}$ is the {\em interface} value of the flux. 

The solution then follows our usual MHD method: we calculate the coefficients, perform a reconstruction step to estimate quantities ``at the face,'' replace the flux with the solution to a Riemann problem (RP), and use this to update the conserved quantities. We will explain this in more detail below.

Note that Eq.~\ref{eqn:faces} is manifestly antisymmetric, so conserved quantities $(V{\bf U})$ are manifestly conserved as desired. 

Additionally, we note that the MFM and MFV methods from \paperone\ differ only at second order in how they handle advection between particles. In diffusion-only problems, they are manifestly identical. Therefore, we do not show both (although we have confirmed their identical results), but will, for simplicity, adopt MFM as our reference method.

\vspace{-0.5cm}
\subsection{Gradient Estimation \&\ Reconstruction}

We require gradients for all quantities; to do this, we adopt the standard gradient estimator in {\small GIZMO}, a moving, second-order accurate and consistent least-squares estimator. For a scalar $f$, this is 
\begin{align}
\label{eqn:gradient} (\nabla f)_{i}^{\alpha} &= \sum_{j}(f_{j} - f_{i})\,\left({\bf W}_{i}^{-1}\right)^{\alpha\beta}\,({\bf x}_{j}-{\bf x}_{i})^{\beta}\,\omega_{j}({\bf x}_{i}) \\
\label{eqn:gradient.matrix} {\bf W}_{i}^{\alpha\beta} &\equiv \sum_{j}\,({\bf x}_{j}-{\bf x}_{i})^{\alpha}\,({\bf x}_{j}-{\bf x}_{i})^{\beta}\,\omega_{j}({\bf x}_{i})
\end{align}
here we assume an Einstein summation convention over the indices $\beta$ corresponding to the spatial dimensions, and $\omega_{j}({\bf x}_{i})$ is an (arbitrary) weight function defined in \paperone. This estimator is second-order accurate for an {\em arbitrary} mesh/particle configuration, minimizes the (weighted) least-squares deviation $\sum_{j}\omega_{j}\,| f_{i} + \nabla f_{i}\cdot ({\bf x}_{j}-{\bf x}_{i}) - f_{j}|^{2}$, and has been applied in a wide range of different numerical methods \citep[see e.g.][]{onate:1996.fpm,kuhnert:2003.finite.pointset.method,maron:2003.gradient.particle.mhd,luo:2008.compressible.flow.galerkin,lanson.vila:2008.meshfree.consistency}. Note that Eq.~\ref{eqn:gradient} applies to a scalar field $f$. For a general tensor ${\bf q}$, we apply Eq.~\ref{eqn:gradient} separately to every component $f = q_{\alpha\beta\gamma...}$ to determine all partial derivatives needed to construct $\nabla \otimes {\bf q}$ and ${\bf F}_{\rm diff}$. The gradients are slope-limited as described in \paperone, such that they do not create new local extrema within the interacting kernel.

In the reconstruction step we must extrapolate the values from $i$ and $j$ to the left and right sides of the face. For hydrodynamics and MHD, we perform a linear reconstruction of all the MHD quantities based on their slope-limited gradients (see \paperone). For additional quantities needed for diffusion, our default approach is a first-order reconstruction, e.g.\ $(\nabla\otimes{\bf q})_{{R,\,L}} = (\nabla\otimes{\bf q})_{{i,\,j}}$. This is easy to implement, and most stable, but comes at the cost of greater numerical diffusion. We have therefore also considered limited tests using the ``double linear'' reconstruction of \citet{munoz:2013.viscous.flows.arepo}. This amounts to treating $\nabla\otimes{\bf q}$ like any other primitive variable. In a first loop, we calculate $\nabla\otimes{\bf q}$ with our standard method; in a second loop, we calculate $\nabla \otimes (\nabla \otimes {\bf q})$, and use this to linearly reconstruct $(\nabla\otimes{\bf q})_{{R,\,L}}$ (or any component of it) like any other primitive variable. This gives a simple, second-order Taylor-series representation which implicitly includes an appropriately larger stencil (since the second pass sums over the values $(\nabla \otimes {\bf q})_{j}$, themselves constructed from the ${\bf q}_{k}$ in all neighbors of $j$). To ensure smoothness, the reconstructed gradients in the double-linear method are slope-limited with respect to the particle-centered gradients using the same limiter we adopt for all the usual MHD quantities (see \paperone).


\vspace{-0.5cm}
\subsection{The Riemann Problem}
\label{sec:riemann}

We treat diffusion in operator-split fashion from the pure-MHD RP. We have experimented with several Riemann solvers for the diffusion RP, and find the best compromise between numerical diffusion, stability, and flexibility using the \citet{Harten:1983:hll.riemann.solver} (HLL) solver in the Lagrangian frame. In this case: 
\begin{align}
\label{eqn:Fdiff.hll}
{\bf F}_{{\rm diff},\,ij}^{\ast} = \frac{\lambda^{+}\,{\bf F}_{{\rm diff},\,L}  
- \lambda^{-}\,{\bf F}_{{\rm diff},\,R}  
+ \alpha\,\lambda^{+}\lambda^{-}\,({\bf U}_{R}-{\bf U}_{L})}{\lambda^{+} - \lambda^{-}}
\end{align}
where the maximum/minimum wavespeeds $\lambda^{+}$/$\lambda^{-}$ are determined appropriate to the problem. Here ${\bf U}_{R}$, ${\bf U}_{L}$ are the appropriate right/left states reconstructed at the face following \paperone\ (extrapolated from the particle-centered values to the face with the derivatives of ${\bf U}$, with a MINMOD slope limiter).

Consider the case where we are using our second-order reconstruction in a Lagrangian frame. We adopt the wavespeed estimate from \paperone, $\lambda^{+} = {\rm MAX}(v_{L},\,v_{R}) + c_{\rm fast}$, $\lambda^{-} = {\rm MIN}(v_{L},\,v_{R}) - c_{\rm fast}$. Here $c_{\rm fast}$ is a fastest wavespeed determined by the 1D RP. Then, because in our default {\small GIZMO} method our frame is moving with $v= (v_{L}+v_{R})/2$, in the Lagrangian frame we simply obtain $\lambda^{+}=-\lambda^{-} = |v_{R}-v_{L}|/2 + c_{\rm fast}$, in which case the HLL solution reduces to the Global Lax Friedrich (GLF) function: 
${\bf F}_{{\rm diff},\,ij}^{\ast} = ({\bf F}_{{\rm diff},\,L} +{\bf F}_{{\rm diff},\,R})/2 -  \alpha\,(|v_{R}-v_{L}|/2 + c_{\rm fast})\,({\bf U}_{R}-{\bf U}_{L})/2$. 

Although using the GLF flux in {\em non}-Lagrangian methods tends to be excessively diffusive, here it gives nearly indistinguishable results from using the HLL solution with another wavespeed estimate (e.g.\ Roe wavespeeds or the exact eigenvalues of the Jacobian $\partial {\bf F}/\partial {\bf U}$). This is because the first-order term owing to frame motion is automatically accounted for by the Lagrangian nature of the method. Even if we use a double-linear reconstruction, we see only a tiny (percent-level) increase in diffusion using this particular form for the RP solution.

Using the above with $\alpha=1$ gives an effective numerical diffusivity $\kappa_{n}\sim c_{s}\,|{\bf x}|_{ij}$, which can easily exceed the physical $\kappa_{p} \sim \| {\bf K} \|$ by large factors at low resolution. We prevent this by replacing Eq.~\ref{eqn:Fdiff.hll} with the flux-limited equation: 
\begin{align}
\label{eqn:fhll.1} {\bf F}_{\rm HLL} &= {\rm MINMOD}\left( (1+\psi)\,{\bf F}_{\rm 2},\, {\bf F}_{\rm 2} + {\bf F}_{\bf U} \right) \\
{\bf F}_{\rm 2} &= \frac{\lambda^{+}\,{\bf F}_{{\rm diff},\,L} - \lambda^{-}\,{\bf F}_{{\rm diff},\,R}}{\lambda^{+} - \lambda^{-}} \\
{\bf F}_{\bf U} &\equiv \alpha\,\left[ \frac{\lambda^{+}\lambda^{-}\,({\bf U}_{R}-{\bf U}_{L})}{\lambda^{+} - \lambda^{-}}\right] \\ 
\label{eqn:alpha} \alpha &\equiv \left( \frac{ \| {\bf K}^{\ast}\cdot (\nabla\otimes{\bf q})^{\ast} \|}{ \| {\bf K}^{\ast} \|\,\| (\nabla\otimes{\bf q})^{\ast} \| }  \right)\,\left( \frac{0.2 + r}{0.2 + r + r^{2}} \right) \\ 
\label{eqn:fhll.2} r &\equiv \left( \frac{\left| v_{R}-v_{L} \right|}{2} + c_{{\rm fast}}^{\ast}\right)\, \frac{\| {\bf x}_{j}-{\bf x}_{i} \| \, \| {\bf q}^{\ast} \|}{\| {\bf K}^{\ast} \| \, \| {\bf U}^{\ast} \|} 
\end{align}
where above ${\bf F}_{2}$ is the flux we would obtain in the simplest 2nd-order accurate (but numerically unstable) formulation.  ${\bf F}_{\bf U}$ is the diffusive flux from the Riemann problem, appropriately limited by the function $\alpha$. Here $\| {\bf x} \|$ refers to the Frobenius norm of the tensor ${\bf x}$, and $f^{\ast}$ refers to the interface value of the primitive variable $f$; where possible we adopt these from the appropriate solution to the pure-MHD RP, otherwise we approximate $f^{\ast} = (f_{L}+f_{R})/2$. 

The first term in $\alpha$ (in ${\bf K}$ and $\nabla\otimes{\bf q}$) limits ${\bf F}_{\bf U}$ for the anisotropic case, vanishing as $F_{\rm diff}$ does where there is full anisotropic suppression (even at low resolution). Note that the form of ${\bf K}\cdot(\nabla\otimes {\bf q})$ in Eq.~\ref{eqn:alpha} follows directly from the form of ${\bf F}_{\rm diff} = {\bf K}\cdot(\nabla\otimes {\bf q})$. For the cases where $F_{\rm diff}$ is represented by a different linear combination of gradients, the term in Eq.~\ref{eqn:alpha} should be modified accordingly.
 
The second term in $\alpha$ ensures $\alpha\rightarrow1$ at small $r$ and $\alpha\rightarrow1/r$ at large $r$ \citep[the functional form is motivated by][]{rosdahl:m1.method.ramses}; together with the MINMOD application this prevents the numerical diffusivity from exceeding physical values by more than some tolerance parameter $\psi$. As usual this parameter represents some tradeoff: increasing $\psi$ gives smoother solutions at the expense of numerical diffusivity. We find our qualitative results are robust for all $0.05 < \psi < 1$, and use $\psi=0.1$ as our default.

Finally, we also compute a ``direct'' flux based on pairwise direct-difference gradients, and use this to restrict ${\bf F}_{{\rm diff},\,ij}^{\ast}$ via:
\begin{align}
({\bf F}\cdot {\bf A})_{\rm direct} &\equiv -\left[ \frac{{\bf K}_{i}+{\bf K}_{j}}{2}\cdot(\nabla\otimes{\bf q})_{\rm dir}\right]
\cdot \|{\bf A}_{ij} \|\,\frac{{\bf x}_{j}-{\bf x}_{i}}{\|{\bf x}_{j}-{\bf x}_{i} \|} 
 \\ 
(\nabla\otimes{\bf q})_{\rm dir} &\equiv \frac{({\bf x}_{j}-{\bf x}_{i})\otimes({\bf q}_{j}-{\bf q}_{i})}{\|{\bf x}_{j}-{\bf x}_{i} \|^{2}} \\ 
\label{eqn:direct.limiter} {\bf F}_{{\rm diff},\,ij}^{\ast} &= 
\begin{cases}
      {\displaystyle  {\bf 0}}\ \ \ \ \ \ \ \hfill{\tiny ( {\rm SIGN}[({\bf F}\cdot{\bf A})_{\rm direct}] \ne {\rm SIGN}[{\bf F}_{\rm HLL}\cdot {\bf A}_{ij}]} \\ 
      {\displaystyle}\ \ \ \ \ \ \ \hfill{\tiny {\rm and}\ \ \|({\bf F}\cdot{\bf A})_{\rm direct}\| > \epsilon\, \|{\bf F}_{\rm HLL}\cdot {\bf A}_{ij}\|  ) } \\ 
      {\displaystyle {\bf F}_{\rm HLL}}\ \ \ \ \ \hfill {\tiny {\rm otherwise}} \\ 
\end{cases}
\end{align}
with the tolerance parameter $0\le \epsilon \le 1$. We find stable (albeit slightly more noisy) results in all our problems using values as large as $\epsilon=2$, in fact, and larger values do give improved performance at low resolution on multi-dimensional tests (e.g.\ the diffusing ring), but we adopt $\epsilon=1/2$ as our default here.

%

 
We note that this is not the only way to stabilize the anisotropic diffusion equations. For example, \citet{sharma.2007:anisotropic.diffusion.stability.limiters} propose an elegant slope-limiting method; however, it is not obvious how to extend this to unstructured meshes. Recently, \citet{kannan.2015:anisotropic.conduction.arepo} implemented the method of \citet{gao.wu.2013:anisotropic.diffusion.stable.scheme.arbitrary.meshes} for moving-mesh codes. The advantage of their method is that it is extremum-preserving and generalizes relatively easily to implicit integrators. However, we choose to explore alternatives for three reasons. First, the \citet{gao.wu.2013:anisotropic.diffusion.stable.scheme.arbitrary.meshes} method implicitly uses the lower-order ``direct'' gradients, as opposed to our (in principle arbitrarily high-order) matrix-based gradient estimator, necessarily making the method lower-order. Moreover in flows where there is a clear mean-field gradient but large resolution-scale noise, matrix-based estimators are significantly more robust \citep[see][]{garciasenz:2012.integral.sph,mocz:2014.galerkin.arepo,pakmor.2016:improving.arepo.convergence}. Second, our method here allows for non-linear flux-limiting terms (e.g.\ $\alpha$ above) which allow us to limit numerical diffusion to physical values even at arbitrarily low resolution (potentially important in multi-physics problems where the diffusion may, in some places, dominate only on un-resolved scales). And third, our method here, unlike most in the literature, allows for any (arbitrarily complex) tensor ${\bf q}$, and/or linear combinations of ${\bf K}$ and the elements of $\nabla \otimes {\bf q}$ in the fluxes (relevant for e.g.\ radiation transport, Braginskii viscosity, and the Hall effect).

\vspace{-0.5cm}
\subsection{Timestepping}

In addition to the usual timestep limiters (e.g., the CFL condition, gravitational acceleration-based limiters) which always apply, ensuring numerical stability in explicit methods for diffusion equations requires an additional timestep criterion:
\begin{align}
\label{eqn:timesteps} \Delta t \le \frac{1}{\|{\bf K}\|}\, \left[ \frac{ \| \nabla \otimes {\bf q} \| }{\| {\bf q} \|} + \frac{1}{\Delta x} \right]^{-2}\,\frac{\| {\bf U} \|}{ \| {\bf q} \|}
\end{align}
For the simplest diffusion example, $\partial U/\partial t = \kappa\,\partial^{2} U / \partial x^{2}$ and an appropriately slope-limited gradient, this reduces to the common expression $\Delta t \le \Delta x^{2}/\kappa$. We also require that the inter-particle flux between a particle and any neighbors of the conserved $V{\bf U}$ cannot exceed half the minimum of the $|V{\bf U}|$ in the pair, in a single timestep, but find this criterion is always satisfied if the above (more strict) criterion is as well.

\vspace{-0.5cm}
\section{Examples Implemented}
\label{sec:examples}

We implement this method in the code {\small GIZMO}. A number of specific physical cases have been implemented; our focus here is not on the microphysics but on the numerical methods. Still, it is useful to list the relevant examples, both to explicitly see how they correspond to the general formulation above, and to give examples of the different forms of anisotropy that are physically expected.

\vspace{-0.5cm}
\subsubsection{Isotropic Passive-Scalar Diffusion}

For this simple case, we have
\begin{align}
{\bf K} &= \kappa\,{\bf I} \\ 
{\bf q} &= {\bf U} = n_{\rm scalar}
\end{align}
where $\kappa$ is the (arbitrary) diffusion coefficient, ${\bf I}$ the identity matrix, and $n_{\rm scalar}$ the scalar density.

\vspace{-0.5cm}
\subsubsection{Anisotropic Thermal Conduction}

Isotropic and anisotropic thermal conduction are represented by:
\begin{align}
{\bf K} &= \kappa_{\bot}\,{\bf I} + \kappa_{\|}\,\hat{B}\otimes{\hat{B}} \\ 
{\bf q} &= T   \ \ \ , \ \ \   {\bf U} = \rho\,u
\end{align}
where $T$ is the gas temperature, $\rho$ the gas density and $u$ the internal energy per unit mass, $\kappa_{\bot}$ the isotropic diffusion coefficient, $\kappa_{\|}$ the anisotropic (parallel) diffusion coefficient, and $\hat{B} \equiv {\bf B}/|{\bf B}|$ the direction of the local 
magnetic field. Here we allow either arbitrarily specified $\kappa$, or default to the Spitzer conductivity (for $\kappa_{\|}$, with $\kappa_{\bot}=0$) with a limiter for cases where 
the implied heat transport rate exceeds the free electron streaming rate.

\vspace{-0.5cm}
\subsubsection{Anisotropic Cosmic Ray Diffusion \&\ Streaming}

Cosmic rays diffuse along magnetic field lines. Moreover, as shown by \citet{uhlig:2012.cosmic.ray.streaming.winds}, cosmic-ray streaming can be represented numerically 
via a diffusion operator. Therefore we have, for the simple case of a single-species cosmic-ray population
\begin{align}
{\bf K} &= (\kappa_{\rm diff} + \kappa_{\rm stream})\,\hat{B}\otimes{\hat{B}} \\ 
{\bf q} &= P_{\rm cr} \ \ \ , \ \ \ {\bf U} = e_{\rm cr}
\end{align}
where $P_{\rm cr}$ is the cosmic ray pressure (proportional for a single-species model to the CR number density), $e_{\rm cr}\approx P_{\rm cr}/(\gamma_{\rm cr}-1) = \rho\,u_{\rm cr}$ is the cosmic ray energy density, and $\kappa_{\rm diff}$ and $\kappa_{\rm stream}$ are the appropriate ``effective'' coefficients for diffusion and streaming processes, respectively.

\vspace{-0.5cm}
\subsubsection{Turbulent Eddy Diffusion Models}

In popular ``sub-grid'' models for turbulence, the effect of unresolved eddies is treated as a diffusion process. Following the  \citet{smagorinsky.1963:eddy.approximation.for.diffusion.terms} model, we model this for e.g.\ diffusion of a scalar field via 
\begin{align}
{\bf K} &= \rho\,(C\,\Delta x)^{2}\,\| {\bf S} \|\, {\bf I}\\
{\bf q} &= u_{\rm scalar} = \frac{n_{\rm scalar}}{\rho}   \ \ \ , \ \ \   {\bf U} = n_{\rm scalar}
\end{align}
where $\Delta x$ is the resolution scale (for our meshless methods, defined to be equal to the rms inter-element spacing inside the kernel), $C\sim0.1$ is a constant calibrated to numerical simulations in \citet{smagorinsky.1963:eddy.approximation.for.diffusion.terms}, and ${\bf S}$ is the symmetric shear tensor. 

\vspace{-0.5cm}
\subsubsection{Anisotropic Radiation Transport in the Diffusion Limit}

In the optically thick limit, the radiative transfer moment equations are commonly expressed as a diffusion equation. It is convenient to represent this in the form
\begin{align}
{\bf F}_{\rm diff} &= -{\bf K}\cdot\left(\nabla \cdot {\bf q}\right) \\ 
{\bf K} &= \frac{\lambda\,c}{\kappa_{\nu}\,\rho}\,{\bf I} \\ 
{\bf q} &= n_{\nu}\,\mathbb{D}_{\nu}  \ \ \ , \ \ \  {\bf U} = n_{\nu}
\end{align}
where $c$ is the speed of light, $\kappa_{\nu}$ the opacity at frequency $\nu$, $n_{\nu}$ the photon number (or energy) density, $\lambda$ an optional ``radiative flux limiter,'' and $\mathbb{D}_{\nu}$ the dimensionless Eddington tensor. 
Note the $\nabla \cdot {\bf q}$ instead of $\nabla \otimes {\bf q}$ has no effect on our methodology, it simple re-orders the gradient terms in ${\bf F}_{\rm diff}$. Various numerical methods (e.g.\ flux-limited diffusion and ``variable Eddington tensor'' methods) rely on this description. 

\vspace{-0.5cm}
\subsubsection{Anisotropic Viscosity}
\label{sec:viscosity}

For a viscous fluid, we have
\begin{align}
{\bf q} &= {\bf v} \ \ \ , \ \ \ {\bf U} = \rho\,{\bf v}
\end{align}
where ${\bf v}$ is the velocity and $\rho$ the density. In this case the form of ${\bf F}_{\rm diff}$ depends on the viscous parameterization. In the case of a magnetized fluid, the Braginskii viscosity can be written 
\begin{align}
{\bf F}_{\rm diff} &= -{\bf K}\left[ \hat{\bf K}:\left(\nabla \otimes {\bf q}\right)\right] \\ 
{\bf K} &= 3\nu_{\|}\,\left(\hat{B}\otimes\hat{B} - \frac{1}{3}\,{\bf I} \right)\\
\hat{{\bf K}} &= \frac{ {\bf K}}{3\,\nu_{\|}} = \hat{B}\otimes\hat{B} - \frac{1}{3}\,{\bf I}
\end{align}
while for the un-magnetized case, it is common practice to decompose the viscosity into shear ($\eta$) and bulk ($\zeta$) terms, following
\begin{align}
{\bf F}_{\rm diff} &= {\bf \Pi}_{\eta} + {\bf \Pi}_{\zeta}\\
{\bf \Pi}_{\eta} &= -{\bf K}_{\eta}\left[ \nabla\otimes{\bf q} + (\nabla \otimes {\bf q})^{T} - \frac{2}{3}\,(\nabla\cdot{\bf q}) \right] \ , \ \ {\bf K}_{\eta} = \eta\,{\bf I} \\
{\bf \Pi}_{\zeta} &= -{\bf K}_{\zeta}\,(\nabla\cdot{\bf q})  \ \ \ , \ \ \ {\bf K}_{\zeta} = \zeta\,{\bf I}
\end{align}
Here ``$:$'' denotes the double-dot-product. As above, the particular arrangement of ${\bf K}$ and $\hat{\bf K}$ within the double-dot-product (for the magnetized case) or decomposition of $\nabla\otimes{\bf v}$ into shear/bulk terms have no effect on our methodology; they simply re-order the gradients of ${\bf q}$ within ${\bf F}_{\rm diff}$. As with conduction we allow either freely-specified $\eta$, $\zeta$, $\nu_{\|}$, or can calculate the coefficients accoding to Spitzer-Braginskii theory with the appropriate limiters. We also add the corresponding viscous term to the energy equation.

\vspace{-0.5cm}
\subsubsection{Non-Ideal MHD}
\label{sec:non.ideal.mhd}

Astrophysical non-ideal MHD effects are typically parameterized as Ohmic dissipation (controlled by the resistivity $\eta_{O}$), the Hall effect ($\eta_{H}$) and ambipolar diffusion ($\eta_{A}$). All appear as diffusion operators in the induction equation; if we operator-split the ideal MHD term (already solved in {\small GIZMO}), we have
\begin{align}
\frac{d {\bf B}}{d t} &= -\nabla\times\left[ \eta_{O}\,{\bf J} + \eta_{H}\,\left({\bf J}\times{\hat {B}} \right) - \eta_{A}\,\left( {\bf J} \times {\hat{B}}\right)\times {\hat{B}} \right] 
\end{align}
where ${\bf J} = \nabla \times {\bf B}$. We can, with some elaborate algebra, write this as an equation in $\nabla\cdot {\bf F}_{\rm diff}$, but it is easier to cast this directly into the alternative Godunov form:
\begin{align}
\frac{d}{dt}(V\,{\bf U})_{i} = & - \sum_{j}\, {\bf A}_{ij}\times {{\bf F}}_{{\rm diff},\,ij}^{\ast} \\
{\bf F}_{\rm diff} = &\ -[{\bf K}_{O} + {\bf K}_{H} + {\bf K}_{A}]\cdot (\nabla \times {\bf q}) = -{\bf K}\cdot (\nabla \times {\bf q}) \\ 
{\bf K}_{O} &= \eta_{O}\,{\bf I} \\
{\bf K}_{A} &= \eta_{A}\,\left({\bf I} - \hat{B}\otimes\hat{B} \right) \\
{\bf K}_{H} &= \eta_{H}\,
\left[ \begin{array}{ccc}
0 & \hat{B}_{z} & -\hat{B}_{y} \\
-\hat{B}_{z} & 0 & \hat{B}_{x} \\
\hat{B}_{y} & -\hat{B}_{x} & 0 \end{array} \right]\\
{\bf q} &= {\bf B} \ \ \ , \ \ \  {\bf U} = {\bf B}
\end{align}
where $\hat{B}_{x}$, $\hat{B}_{y}$, $\hat{B}_{z}$ are the components of $\hat{B}$ and the coefficients are calculated from the plasma properties. The appropriate magnetic terms are added to the energy equation.

\begin{figure}
\plotonesize{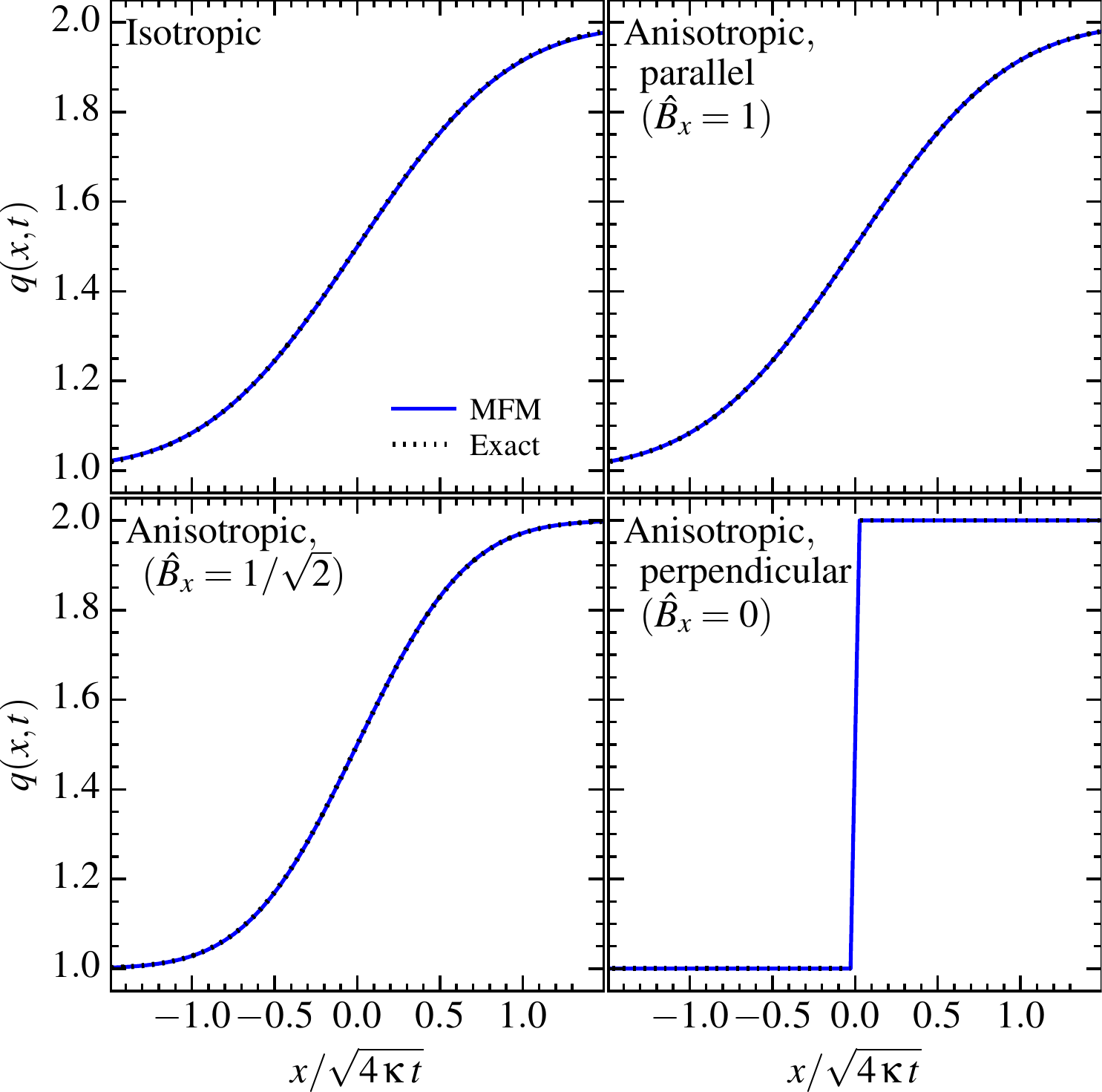}{0.99}
    \vspace{-0.25cm}
    \caption{Diffusing sheet test (\S~\ref{sec:sheet}). A 3D box ($N=64$ elements across plotted $x$-range) is initialized with a step-function discontinuity ($x=0$) in diffusing quantity ${\bf q}$. The solution is scale-free in units shown, and independent of the physics (we confirm identical behavior for implementations of passive-scale diffusion, conduction, cosmic rays, turbulent eddy diffusion, radiation transport in the diffusion limit, viscosity, and Ohmic resistivity; \S~\ref{sec:examples}). We consider isotropic and anisotropic cases with diffusion tensor ${\bf K}=K\,(\hat{B}\otimes\hat{B})$, with parallel, partially-aligned, and perpendicular fields ${\bf B}$. In each case we compare our mesh-free finite-element methods (MFM, our MFV method is manifestly identical on these problems; \S~\ref{sec:mfm}) to exact solutions. 
        \vspacerpostplot 
    \label{fig:diff.sheet.basic}}
\end{figure}

\vspace{-0.5cm}
\section{Numerical Tests}

\subsection{Diffusing Sheet}
\label{sec:sheet}

For a first test in Fig.~\ref{fig:diff.sheet.basic}, we study a simple discontinuity in one dimension, ${\bf q} = {\bf q}_{L}$ for $x<0$, ${\bf q}={\bf q}_{R}$ for $x>0$, with all other properties fixed. To focus on the diffusion equations, we turn off all MHD forces (for now). We also simulate this in a 3D box (with an arbitrarily large periodic domain), because although it is essentially a 1D problem, sensitivity to particle arrangement makes accurate solution much more challenging in 2D or 3D. Finally, we also add noise (which is critical to test for numerical instability): we add a uniform random value to $q$ between $\pm0.05\,q_{L}$, for all particles. 

\vspace{-0.5cm}
\subsubsection{Isotropic Case}

In the isotropic case, after the noise has been (quickly) damped away,\footnote{We have considered both a perfect step-function initial condition, and (alternatively) initializing the profile with the analytic solution corresponding to time $t_{0}$ such that $\sqrt{4\,\kappa\,t_{0}}/\langle \Delta x \rangle \approx 2$. These are indistinguishable after a short time ($\sqrt{4\,\kappa\,t_{0}}/\langle \Delta x \rangle\gtrsim 4$).}  this has the trivial analytic solution
\begin{align}
q(x,\,t) = \frac{q_{R}+q_{L}}{2} + \frac{q_{R}-q_{L}}{2}\,{\rm Erf}\left[ \frac{x}{\sqrt{4\,\kappa\,t}} \right]
\end{align}
The solution is completely self-similar so the absolute values of all quantities are irrelevant; the only numerically important quantity is the number of resolution elements over which the contact discontinuity has been diffused $\sim \sqrt{4\,\kappa\,t}/\Delta x$ (as we evolve the solution, it becomes better-resolved). We therefore plot results at fixed $\sqrt{4\,\kappa\,t}/\langle \Delta x \rangle$.

Our results converge well to the analytic solution with both methods even at low resolution ($\sqrt{4\,\kappa\,t}/\langle \Delta x \rangle\gtrsim 4$). Because the initial noise is grid-scale, it should be damped away on a small timescale $\sqrt{4\,\kappa\,t}/\langle \Delta x \rangle\sim1$; we confirm this.

Note that the solution here is independent of the physics; we have explicitly verified that our implementations of passive-scalar diffusion ($n\propto q(x,\,t)$ above), conduction ($T\propto q$ with ${\bf K}=\kappa_{\bot}{\bf I}$), cosmic rays ($e_{\rm cr}\propto q$, with ${\bf K}=\kappa_{\rm diff}\,{\bf I}$), eddy diffusion ($n\propto q$), isotropic radiation transport ($n_{\nu}\propto q$ with $\mathbb{D}_{\nu}={\bf I}/3$), viscosity (${\bf B}=0$ with $v_{x}=v_{z}=0$, $v_{y}\propto q$, and $\zeta=0$), and Ohmic resistivity ($B_{y}\propto q$) all give {\em identical} results on this test (given the same diffusivity), as they should.

\vspace{-0.5cm}
\subsubsection{Anisotropic Case}

Next we consider the anisotropic case. Here we take ${\bf K}=\kappa\,\hat{B}\otimes\hat{B}$ with $\kappa$ and $\hat{B}$ constant and $|\hat{B}|=1$. In this case,  the solution is identical to the isotropic case but with $\kappa \rightarrow \kappa\,|\hat{B}\cdot\hat{x}|^{2}=\kappa\,\hat{B}_{x}^{2}$, so it is entirely specified by the absolute value of the projection of $\hat{B}$ in the gradient ($\hat{x}$) direction, $|\hat{B}_{x}|$.

We have considered $\sim100$ values between $-1\le \hat{B}_{x} \le 1$ to check for pathological behavior; these are summarized with three representative cases shown here: $|\hat{B}_{x}|=0$ (perpendicular fields, which should completely suppress diffusion), $|\hat{B}_{x}|=1$ (parallel fields; the solution should be identical to the isotropic case), and $|\hat{B}_{x}|=1/\sqrt{2}$. For all scalar diffusion cases (passive scalars, cosmic rays, eddy diffusion, conduction) this gives identical results. For our radiation diffusion, we make the system anisotropic by instead taking $\mathbb{D}=\hat{n}\otimes\hat{n}$, with $\hat{n}=\hat{B}$ constant. We confirm this produces exactly identical solutions to the cases above.

Our MFM method is able to handle all three cases accurately; we confirm that there is zero diffusivity for the perpendicular case (up to machine precision, if our initial ${\bf q}$ depends on $x$ alone) and that the parallel case exactly matches the isotropic case. Convergence is again good even at low resolution $\sqrt{4\,\kappa\,t}/\Delta x \gtrsim 4$, and (we show below) the results are insensitive to noise or particle order.

\begin{figure}
\plotonesize{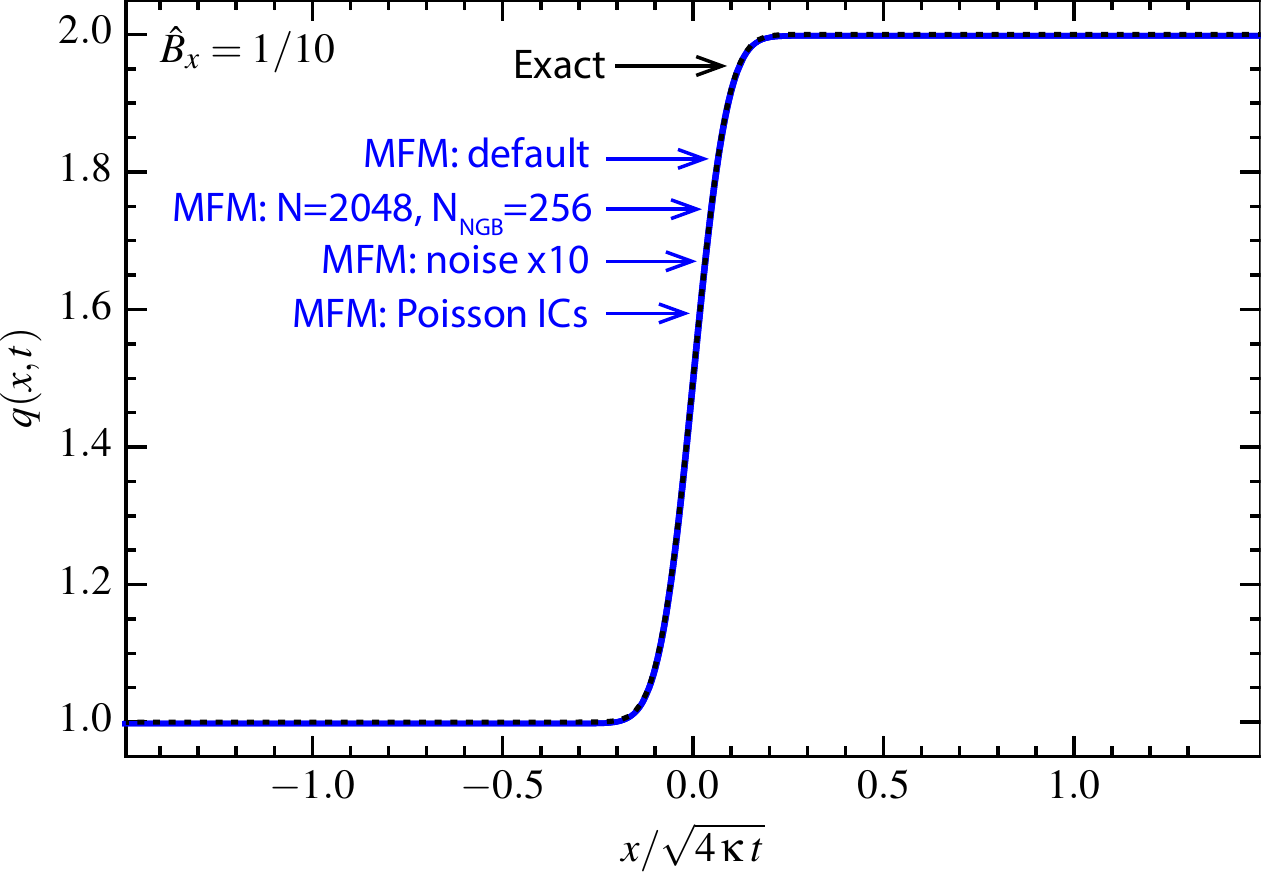}{0.99}
    \vspace{-0.25cm}
    \caption{Anisotropic sheet (as Fig.~\ref{fig:diff.sheet.basic}) with varied ICs (\S~\ref{sec:sheet.variations}). We compare {\bf (1)} A ``default'' case with resolution $N=64$ particles across the plotted range in the $x$-direction ($\approx4$ particles across the range of diffusion at this time, in MFM), and effective neighbor number $N_{\rm NGB}=32$ within the kernel partition function. Particles are initially in a cubic lattice, with noise in ${\bf q}$ equal to $2.5\%$ of the jump value. {\bf (2)} High-resolution ($N=2048$) and neighbor number ($N_{\rm NGB}=256$, with a quintic kernel function). {\bf (3)} Initial noise $=25\%$ of jump value. {\bf (4)} Initial particles laid down randomly (according to a Poisson distribution). Our MFM results are indistinguishable from each other and the exact solution in each case. 
        \vspacerpostplot 
    \label{fig:diff.sheet.mods}}
\end{figure}

\vspace{-0.5cm}
\subsubsection{Dependence on Particle Order, Resolution, Noise, and Neighbor Number}
\label{sec:sheet.variations}

Fig.~\ref{fig:diff.sheet.mods} considers variations of the anisotropic sheet. In order to test whether our methods are sensitive to the local arrangement of particles, we have considered (in both 2D and 3D tests) an initial particle distribution following (1) a regular square lattice, (2) uniformly randomly-distributed particles over the volume (a Poisson distribution), (3) a glass (generated from the random distribution), and (4) a densest sphere packing. In 2D we have also considered a regular triangular and hexagonal grid. For our MFM/MFV method, the results are almost indistinguishable (after the initial noise is damped) in every one of these cases (even the ``worst-case'' Poisson distribution, shown in Fig.~\ref{fig:diff.sheet.mods}), clearly demonstrating that the method does not depend sensitively on particle order. 

We consider resolution tests, varying both the absolute resolution and number of kernel neighbors. Our MFM/MFV results are well-converged even with just $\sim 4-8$ resolution elements across the jump (i.e.\ $(\sqrt{4\,\kappa\,t})/\Delta x\gtrsim4$). The convergence rate of our method is also independent of neighbor number, as shown in \paperone\ and \papertwo\ (going to larger neighbor number simply trades a reduction in noise for increased numerical diffusivity). 

We have also tested for sensitivity to initial noise, varying the seed noise level from $0-25\%$ of the jump level. In all cases our results are stable and do not qualitatively change with respect to this. Likewise, the steepness of the initial discontinuity has no effect on our results (whether we begin with the full solution at a resolved scale, or a perfectly steep discontinuity across a single particle).

\begin{figure}
\plotonesize{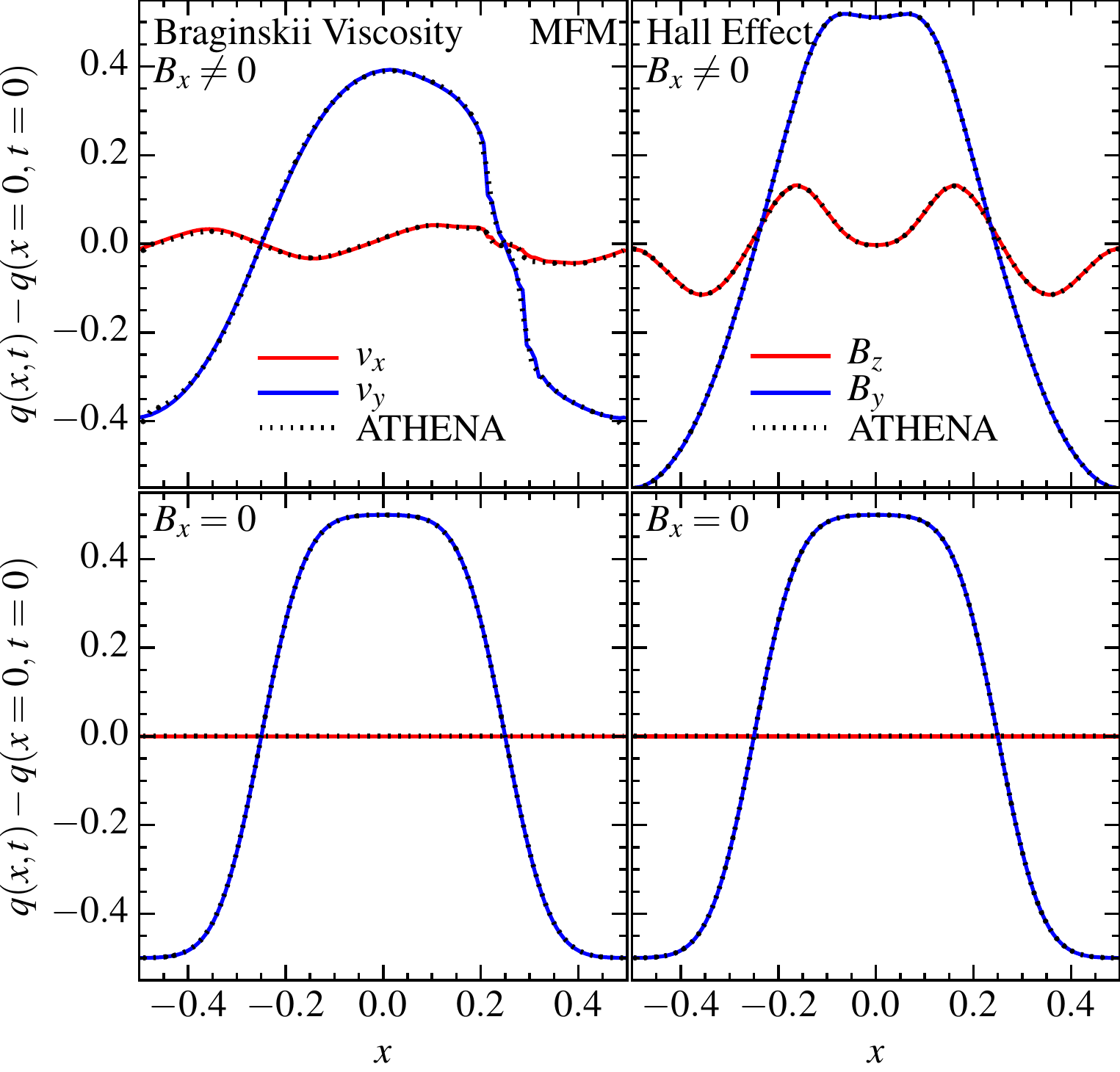}{0.95}
    \vspace{-0.25cm}
    \caption{Anisotropic diffusing sheet problem, as Fig.~\ref{fig:diff.sheet.basic} ($N=64$), but here focusing on two special cases where the solution is not the same as in Fig.~\ref{fig:diff.sheet.basic} (see \S~\ref{sec:diffusion.sheet.special}). There is no analytic solution for the cases shown, so, we compare a converged solution from {\small ATHENA}.
    {\em Left:} Braginskii viscosity (\S~\ref{sec:viscosity}). The initial discontinuity is in $v_{y}$ only, but this form of viscosity generates $v_{x}\ne0$ at later times. We consider two cases: $\hat{B} = (1,\,1,\,0)/\sqrt{2}$ ({\em upper}; this should produce non-zero diffusion) and $\hat{B}=(0,\,1,\,0)$ ({\em lower}; this should have vanishing diffusion).
    {\em Right:} Hall effect (\S~\ref{sec:non.ideal.mhd}). The initial discontinuity is in $B_{y}$ ($B_{x}=$\,constant, $B_{z}$ vanishes). We consider a case which should diffuse ($B_{x}\ne0$; {\em upper}) and one which should be suppressed ($B_{x}=0$; {\em lower}). MFM and {\small ATHENA} agree well in all cases.
        \vspacerpostplot 
    \label{fig:diff.brag}}
\end{figure}

\vspace{-0.5cm}
\subsubsection{Braginskii Viscosity and the Hall Effect}
\label{sec:diffusion.sheet.special}

Two cases in \S~\ref{sec:examples} are more complicated, even in a simple diffusing sheet setup. These are Braginskii viscosity (\S~\ref{sec:viscosity}) and the Hall effect (\S~\ref{sec:non.ideal.mhd}). In both, ${\bf q}$ is a vector (${\bf v}$ or ${\bf B}$, respectively), and even if we initialize a gradient only in one element of that vector and set the other elements to vanish everywhere, the form of the diffusion operator leads to growth of other components of ${\bf q}$. Mathematically, the anisotropic part of the diffusion operator in most cases is a projection operator; here, it also includes a rotation operator. This leads to different non-linear solutions and makes it more challenging to achieve stability. 

Therefore we consider these cases specifically in Fig.~\ref{fig:diff.brag}. Since the non-linear solutions do not have analytic forms even for this simple test, we compare to a converged, high-resolution ($N=2048$ across the plotted domain), one-dimensional solution from the well-tested grid-based code {\small ATHENA} \citep{stone:2008.athena}. We run {\small ATHENA} in its most accurate constrained-transport, PPM-CTU (highest-order) mode. For simplicity, we do not disable the other hydrodynamic forces, but these are not dominant. To avoid certain boundary condition effects in {\small ATHENA} we initialize the test problem in slightly different fashion: we take
\begin{align}
q = 1.5 - 0.5\,\left(1 + {\rm Erf}\left[ \frac{x - 0.25}{0.01} \right] - {\rm Erf}\left[ \frac{x + 0.25}{0.01} \right] \right) 
\end{align}
For the Braginskii problem, we take initial ${\bf v}=(0,\,q,\,0)$ and consider both $\hat{B}=(1,\,0,\,0)$ (which should produce zero diffusion) and $\hat{B}=(1,\,1,\,0)/\sqrt{2}$ (which should diffuse); we take $\rho=u=1$, and $|{\bf B}|=10^{-6}$ to be small so it has no effect on the dynamics except to control the anisotropy. For the Hall problem, we take vanishing initial velocities and ${\bf B}=10^{-6}\,(0,\,q,\,0)$ (no diffusion) ${\bf B}=10^{-6}\,(1,\,q,\,0)$ (diffusion). Our MFM simulations are run in 3D boxes with resolution $N=128$ across the plotted domain.

Fig.~\ref{fig:diff.brag} shows these develop non-zero $v_{x}$ and $B_{z}$, despite these vector components initially vanishing. In both cases our MFM/MFV methods produce solutions in excellent agreement with {\small ATHENA}, even well into non-linear evolution. Both cases with diffusion, and cases with full anisotropic suppression, are captured.

\begin{figure}
 \begin{tabular}{r}
  \includegraphics[width=0.8104\columnwidth]{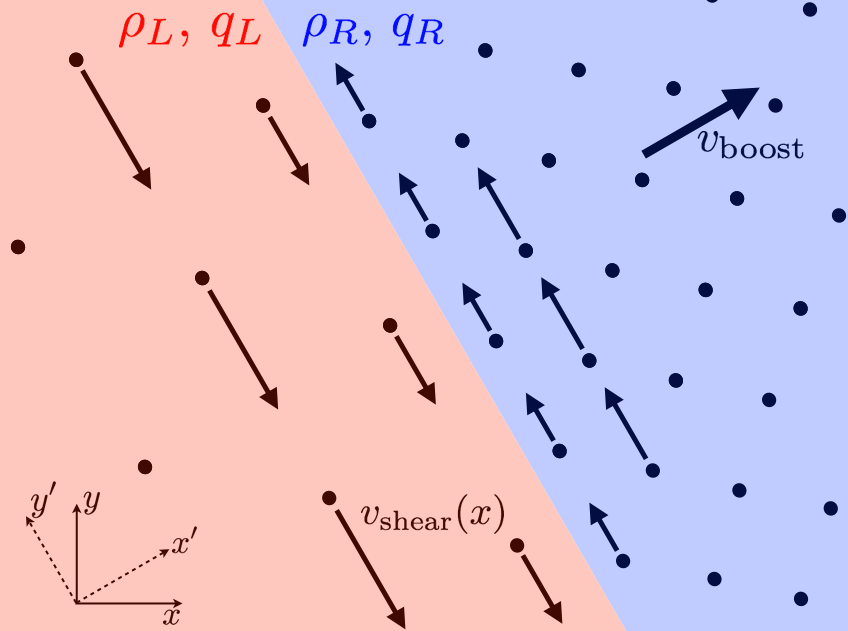} \\
  \includegraphics[width=0.9215\columnwidth]{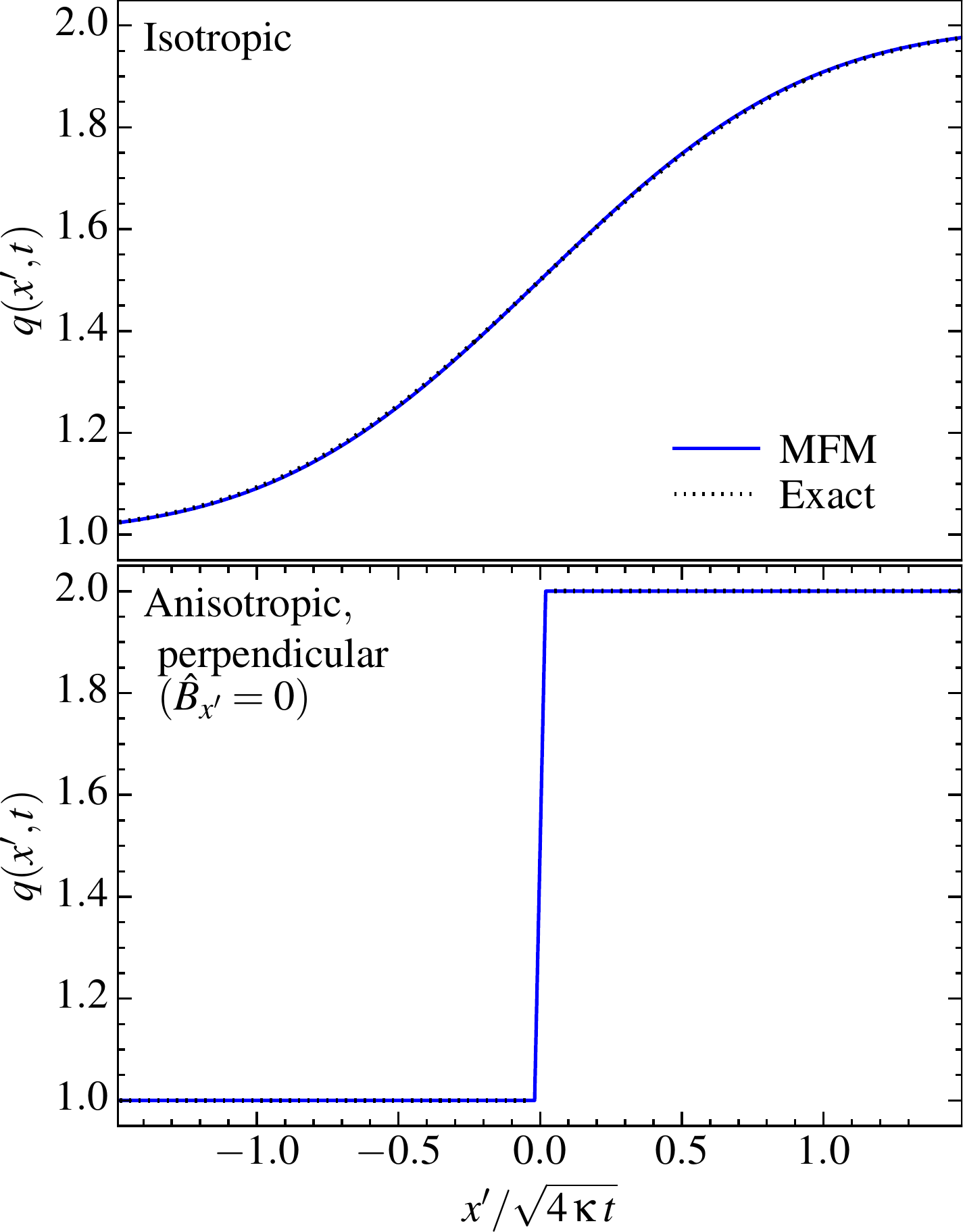} 
 \end{tabular}
    \vspace{-0.25cm}
    \caption{Diffusion across a super-sonically moving, rotated, shearing contact discontinuity (\S~\ref{sec:contact}), as Fig.~\ref{fig:diff.sheet.basic}, with full MHD physics enabled. {\em Top:} Illustration of the initial problem setup. We measure properties along the $x^{\prime}$ axis, perpendicular to the initial discontinuity in density and ${\bf q}$.     
    We consider an isotropic case ({\em middle}) and an anisotropic, perpendicular (fully-suppressed; {\em bottom}) case. The contact discontinuity means the particle arrangement is different on either side of $x^{\prime}=0$, and the shear field means that the arrangement of neighbor particles is constantly changing. These do not destroy our solution.
        \vspacerpostplot 
    \label{fig:diff.sheet.contact}}
\end{figure}

\begin{figure}
\plotonesize{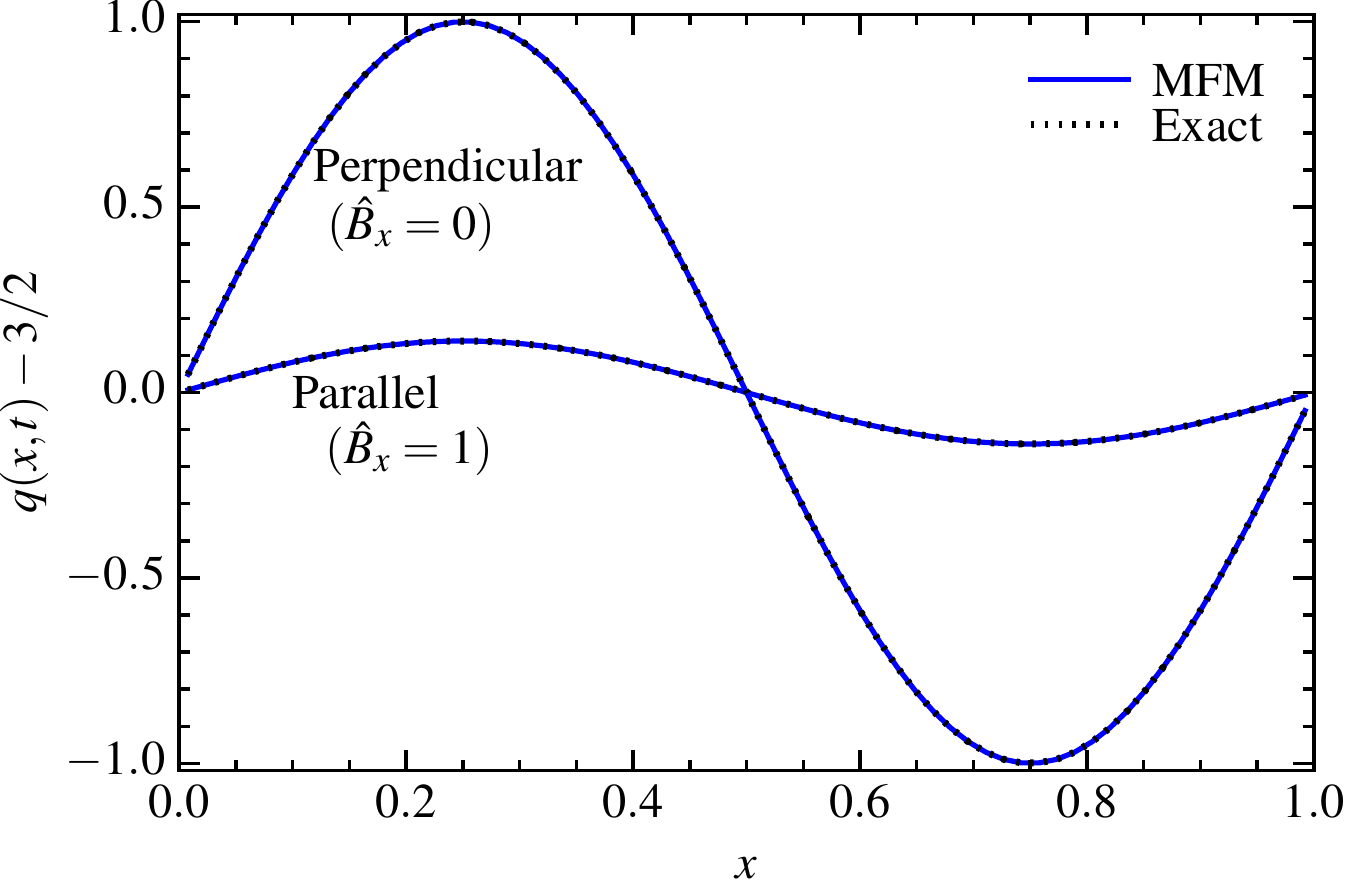}{0.99}
    \vspace{-0.25cm}
    \caption{Diffusion of a sinusoidal pulse (${\bf q} \propto \sin{(2\pi\,x)}$; \S~\ref{sec:sine}), in a 1D periodic box with $0<x<1$ with $N=64$ elements across the $x$-range. We show anisotropic cases with perpendicular (full suppression) and parallel (no suppression) fields. In the perpendicular case, the initial pulse does not evolve; in the parallel case, the amplitude of the pulse decays as $\exp{(-\pi^{2}\,4\,K\,t)}$ (we show results at $4\,K\,t = 1/5$). 
        \vspacerpostplot 
    \label{fig:diff.sine}}
\end{figure}

\vspace{-0.5cm}
\subsection{Diffusion Across a Moving, Rotated, Shearing Contact Discontinuity}
\label{sec:contact}

We next consider a more challenging variation of the diffusion sheet, illustrated in Fig.~\ref{fig:diff.sheet.contact}. We {\bf (1)} re-enable our normal MHD physics. {\bf (2)} Insert a contact discontinuity at $x=0$ (the location of the jump in ${\bf q}$) with a density jump of a factor of $2$, so the diffusion (of some passive scalar) is across the discontinuity. This implies a different particle arrangement (since our particles are equal-mass) across $x=0$. {\bf (3)} Make the discontinuity shearing, with $v_{y} = x/2$. This means the particle geometry around the diffusing interface is constantly being re-arranged. {\bf (4)} Uniformly boost the system by ${\bf v}_{\rm boost}=(10,\,3,\,2)\,c_{s}$. {\bf (5)} Rotate the entire system by $+35^{\circ}$.

None of these changes the physical solution for the diffusion. However they are all, in principle, numerically challenging. Because our methods are fully Lagrangian, our solutions are trivially invariant to the boost {\bf (4)} and rotation {\bf (5)} operations. This is not the case, however, for Eulerian methods. Our MFM method is also invariant to {\bf (1)} and {\bf (2)}, i.e.\ the evolution of a stable contact discontinuity produces vanishing fluxes, and the gradient estimator is explicitly insensitive to particle arrangement within the kernel for linear gradients (this is not always the case in other Lagrangian methods such as SPH, where solutions around contact discontinuities can depend on particle arrangement; see references in \S~\ref{sec:intro}). 

Fig.~\ref{fig:diff.sheet.contact} shows that, despite these complications, our MFM solution still agrees very well with the exact result. There is some noise around $x=0$, introduced by the shear {\bf (3)}, particularly around the contact discontinuity, but it is not visible on the scale plotted. This constant re-arrangement of the particles (hence constant re-arrangement of the effective faces and implicit mesh) introduces ``grid noise'' \citep[for detailed discussion of this noise term see][]{duffell:2014.smooth.moving.mesh.motion,hopkins:gizmo,mocz:2015.mesh.regularization.reduce.moving.mesh.grid.noise,pakmor.2016:improving.arepo.convergence}.

A detailed comparison with grid-based methods is outside the scope of this paper; however, this problem is very challenging for such codes. A moving contact discontinuity, especially one not aligned with the grid, produces large numerical diffusivity at low resolution \citep[see][]{hopkins:gizmo,springel:arepo}. Using {\small ATHENA} to run a version of this problem with anisotropic conduction, we find the numerical diffusivity exceeds the physical diffusion even in the isotropic case unless we use more than $\sim 512$ elements across the plotted range (compared to $64$ here); even higher resolution is required for the anisotropic case.

\begin{figure}
\plotonesize{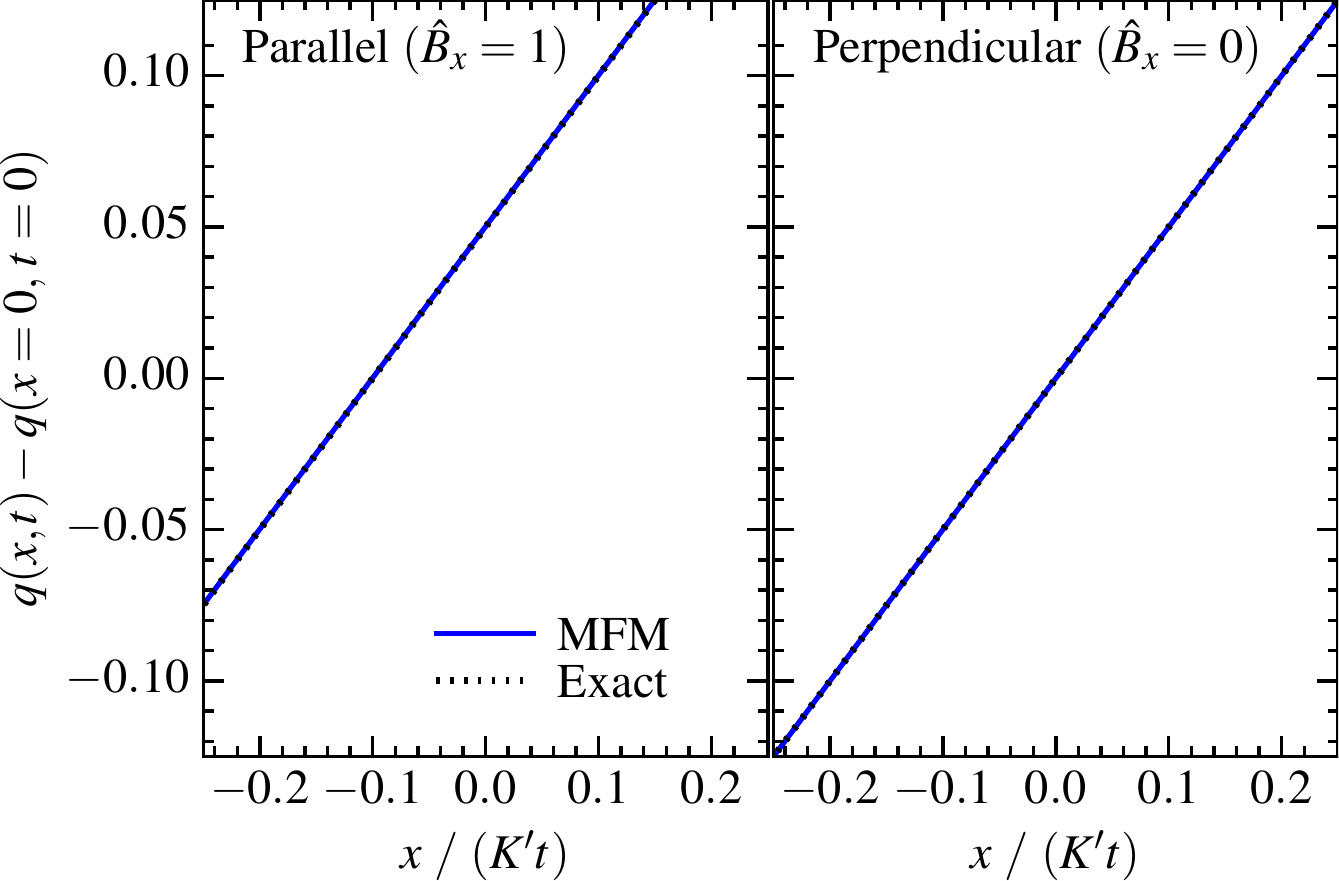}{0.99}
    \vspace{-0.25cm}
    \caption{Diffusion with a variable diffusivity ${\bf K}$ and diffused quantity ${\bf q}$ (\S~\ref{sec:variable}). Both vary linearly in $\hat{x}$. The box is 3D with $N=64$ elements across the $x$-range, but the results are converged (up to tested $N=4096$). We consider anisotropic cases with parallel ({\em left}; no suppression) and perpendicular ({\em right}; full suppression) fields.
        \vspacerpostplot 
    \label{fig:diff.line}}
\end{figure}

\vspace{-0.5cm}
\subsection{Sinusoidal Temperature Distribution}
\label{sec:sine}

Fig.~\ref{fig:diff.sine} considers a one-dimensional test problem in which a scalar $q$ follows a sinusoidal distribution from \citet{arth.2014:anisotropic.conduction.sph.gadget}. This tests the same physics as the later stages of the diffusing sheet, but is much ``easier'' (since it is 1D and there is never a steep gradient). No other MHD physics are active, and we take a periodic box of unit length $L=1$, with unit density and sound speed and $\gamma=5/3$, with the physical solution
\begin{align}
q(t,\,x) = \frac{3}{2} + \sin{\left( 2\pi\,x \right)}\,\exp{\left(-4\pi^{2}\,K\,\hat{B}_{x}^{2}\,t \right)}
\end{align}
initialized at $t=0$. We have confirmed (as expected) that all the conclusions from our diffusing sheet tests apply in this test.


\vspace{-0.5cm}
\subsection{Variable Diffusivity}
\label{sec:variable}

We previously took ${\bf K}$ to be constant. Here, consider the case $q(x,\,t=0) = q_{0}+q^{\prime}\,x$, $K(x,\,t)=K_{0}+K^{\prime}\,x$; this produces the analytic solution $q(x,\,t) = q(x,\,t=0) + q^{\prime}\,K^{\prime}\,t$. In the simple anisotropic case with ${\bf K}=K\, \hat{B}\otimes\hat{B}$ this becomes $q(x,\,t) = q(x,\,t=0) + |\hat{B}|_{x}^{2}\,q^{\prime}\,K^{\prime}\,t$. We consider this in the same 3D box setup as before (here with large enough distance in the $x$-direction so that the boundary conditions do not enter the considered domain).

The results are shown in Fig.~\ref{fig:diff.line}. Our methods perform well in both the anisotropic case and (more trivially, therefore not shown), the isotropic case.

\begin{figure}
\plotonesize{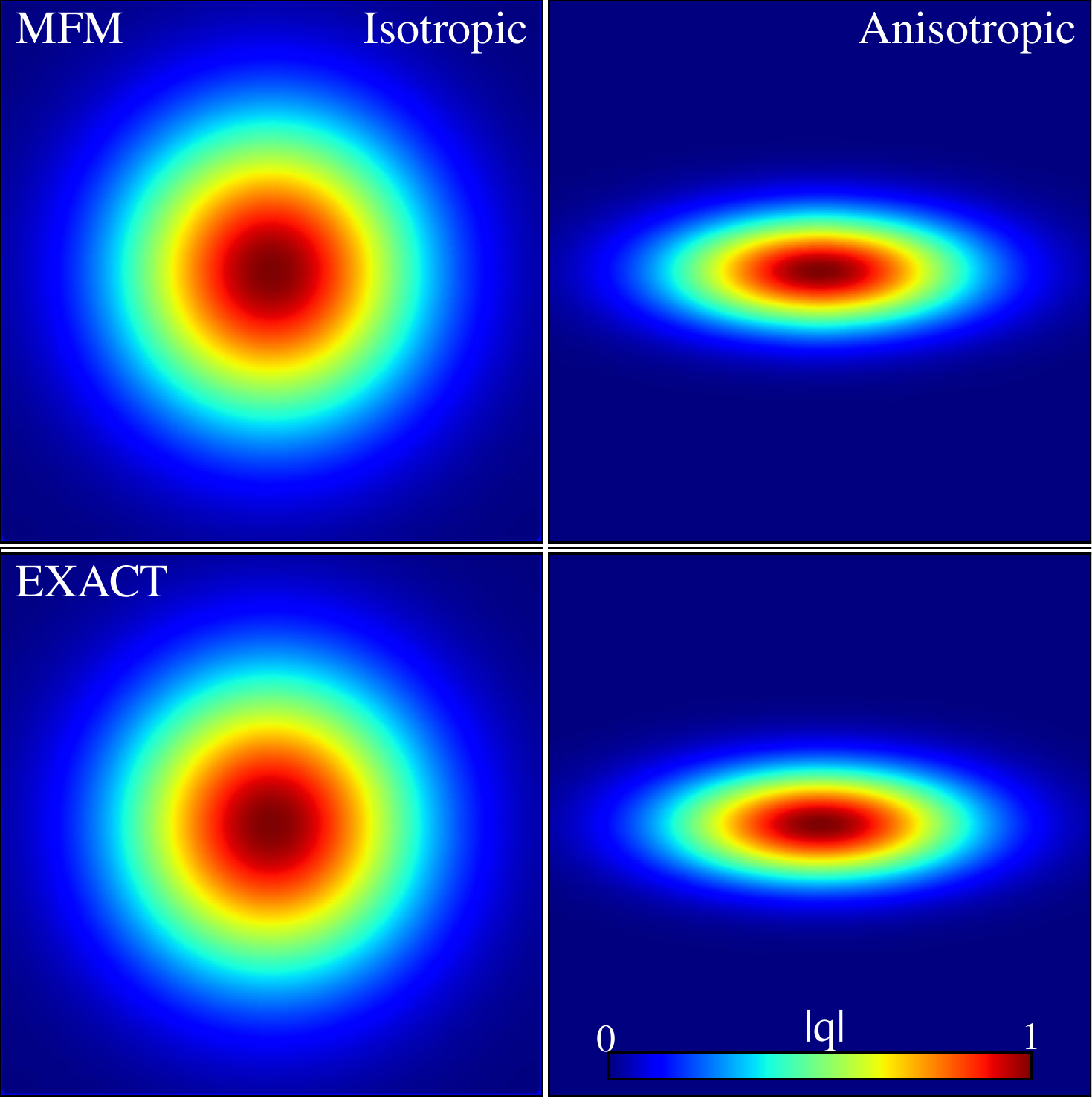}{0.95}
    \vspace{-0.25cm} 
    \caption{Three-dimensional Gaussian pulse test (\S~\ref{sec:pulse}). A Gaussian distribution of the diffused quantity ${\bf q}$ is injected at the center with initial spherically symmetric width $\epsilon=0.05$. We plot a slice through the $x-y$ plane, at time $t=0.8$ for diffusivity $K=0.01$, in a box of unit size with $128^{2}$ elements. Color denotes $|{\bf q}|$, scaled linearly, in arbitrary units (so we simply set ${\rm MAX}[|{\bf q}|]=1$).
    {\em Left:} Isotropic diffusion. The Gaussian simply expands; all methods accurately capture the exact solution ({\em bottom}). {\em Right:} Anisotropic diffusion with ${\bf K}=K\,\hat{B}\otimes\hat{B}$ and $\hat{B}=\hat{x}$. The distribution should expand only in the $x$-direction. MFM captures this, with minor artifacts from the initial particle arrangement which converge away at higher resolution. 
        \vspacerpostplot 
    \label{fig:pulse}}
\end{figure}

\vspace{-0.5cm}
\subsection{Gaussian Pulse}
\label{sec:pulse}

We now consider a multi-dimensional problem -- the diffusion of a quantity injected as a $\delta$-function instantaneously into a homogenous background. In a periodic box of unit size, we initialize a 3D, spherically-symmetric Gaussian for ${\bf q}$ centered on the origin, where the diffused quantity is treated as a passive scalar with all other background properties constant (so no other MHD effects appear). In the isotropic case, this evolves as:
\begin{align}
q({\bf x},\,t) = \frac{q_{0}\,(2\pi)^{-3/2}}{(\epsilon^{2} + 2\,\kappa\,t)^{3/2}} \exp{\left[-\frac{1}{2}\left(\frac{x^{2}+y^{2}+z^{2}}{\epsilon^{2} + 2\,\kappa\,t} \right) \right]}
\end{align}
where $q_{0}$ is an arbitrary normalization, $\kappa$ is the diffusivity, $t$ the time since the problem was initialized, and $\epsilon$ defines the initial width of the distribution ($\epsilon\rightarrow0$ becomes a $\delta$-function; larger $\epsilon$ correspond to starting from an already-evolved solution). We take $\epsilon=0.05$, comparable to our inter-particle spacing, so that there is a well-defined gradient in our initial condition.

In the anisotropic case, if we assume ${\bf K}=K\,\hat{B}\otimes\hat{B}$ with constant (in space and time) $\hat{B}$, we can always define our axes so that $\hat{B}=\hat{x}$; then this evolves as:
\begin{align}
\nonumber q({\bf x},\,t) = \frac{q_{0}\,(2\pi)^{-3/2}}{\epsilon^{2}\,({\epsilon^{2} + 2\,\kappa\,t})^{1/2}} \exp{\left[-\frac{1}{2}\left( \frac{x^{2}}{\epsilon^{2}+2\,\kappa\,t} + \frac{y^{2}+z^{2}}{\epsilon^{2}} \right) \right]}
\end{align}
i.e.\ $q$ diffuses normally along $\hat{B}$, and not perpendicular.

Fig.~\ref{fig:pulse} shows the results. The isotropic case is easily recovered accurately (even at lower resolution than shown). MFM is also able to recover the anisotropic case; however at low resolution (a $64^{3}$ box) the agreement is not perfect, as some artifacts from the grid structure (here particles were initially laid in a Cartesian grid) are present. These are invisible by-eye if we go to $>256^{3}$ resolution, although still measurable in the L1 error norm. In multi-dimensional problems such as this, the perpendicular width of structures must, in general, be a few particles across before complete anisotropy can be fully captured. This is required so that a reliable gradient in the relevant direction can be determined (similar to the requirement in grid-based codes).

\begin{figure}
\plotonesize{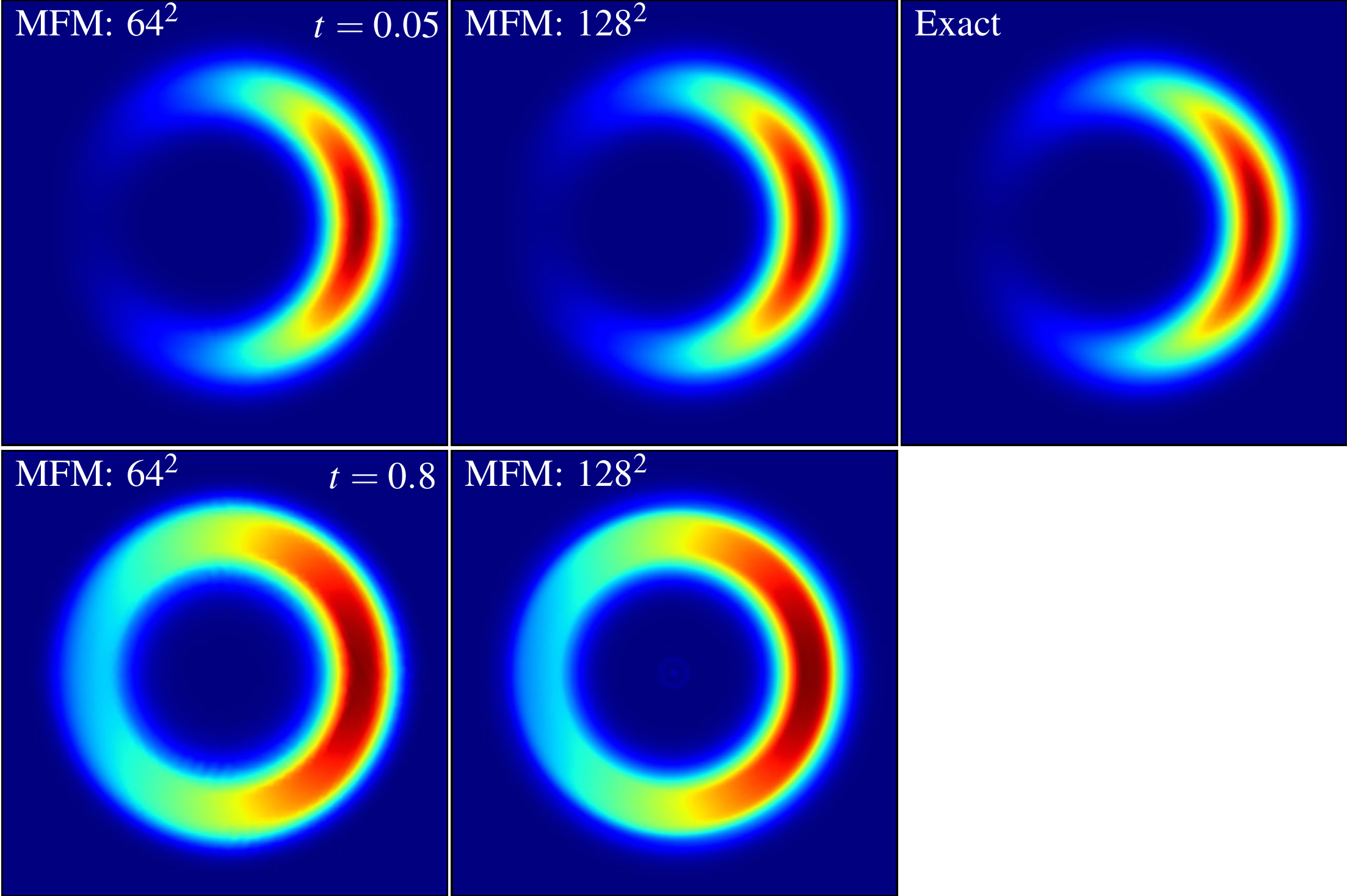}{1.02}
    \vspace{-0.25cm} 
    \caption{The diffusing ring test problem (\S~\ref{sec:ring}; style as Fig.~\ref{fig:pulse}). A ``hot spot'' (high-$|{\bf q}|$) within a small radial annulus is initialized with pure azimuthal fields; this should diffuse into a ring without additional radial diffusion. We compare MFM at two different resolutions (labeled), and both early ({\em top}) and late ({\em bottom}) times. For early times there is an approximate analytic solution, shown; for late times there is no analytic solution, but the high-$|{\bf q}|$ material should remained confined to the same radial annulus as at early times, gradually diffusing around the ring until it is isothermal. In MFM/MFV we see good agreement with the expected behaviors at both times, even at low resolution. There is some numerical radial diffusion, but this gradually converges away as we go towards higher resolution. 
        \vspacerpostplot 
    \label{fig:ring}}
\end{figure}

\begin{figure}
\plotonesize{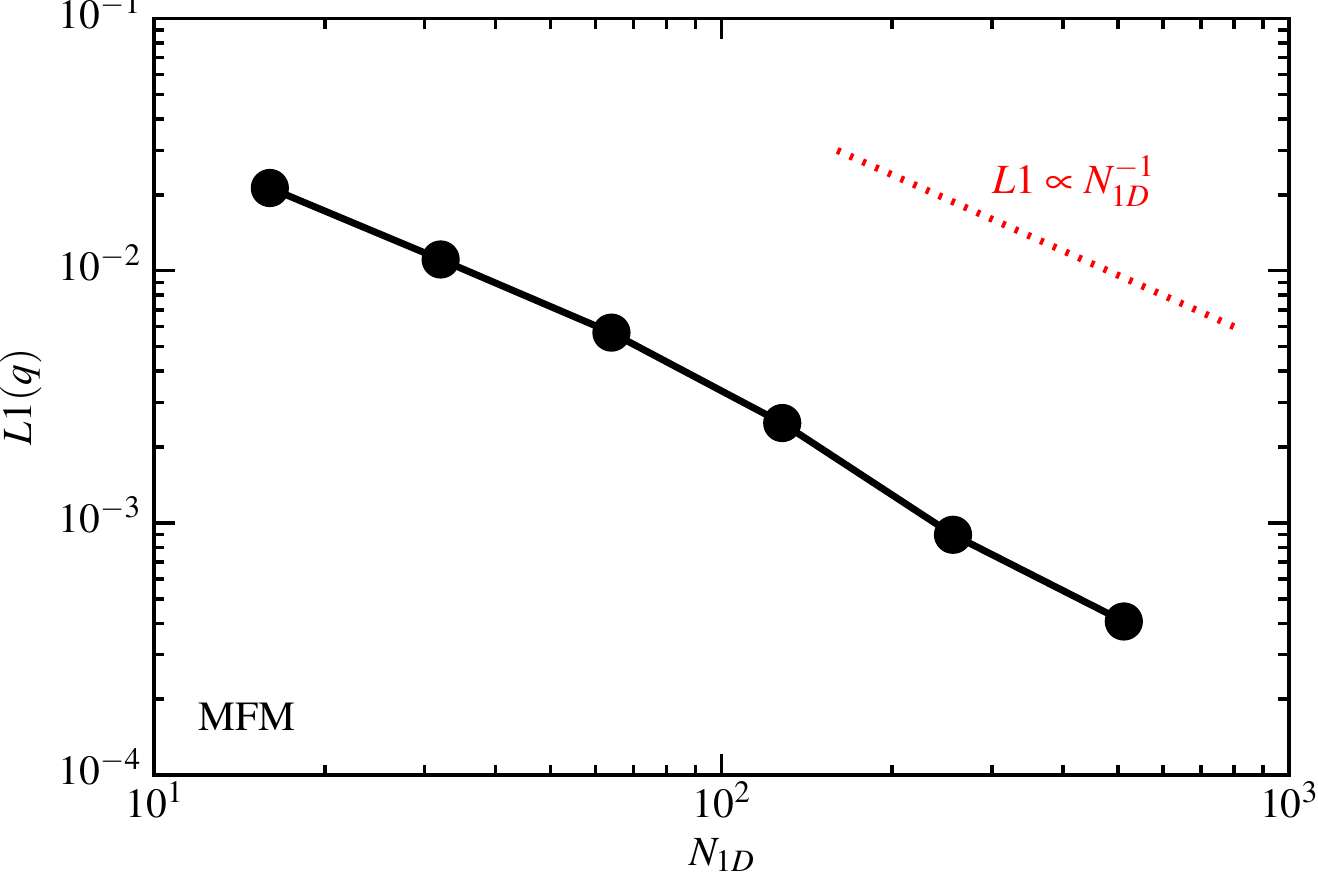}{1.02}
    \vspace{-0.6cm} 
    \caption{Convergence in the diffusing ring test (Fig.~\ref{fig:ring}), with our MFM method. We plot the L1 error norm in $q$ averaged over the domain as a function of the number of elements on a side $N_{1D}$, at a fixed time $t=0.2$. Convergence is close to the ideal linear scaling (dotted red). 
        \vspacerpostplot 
    \label{fig:ring.convergence}}
\end{figure}

\begin{figure}
\plotonesize{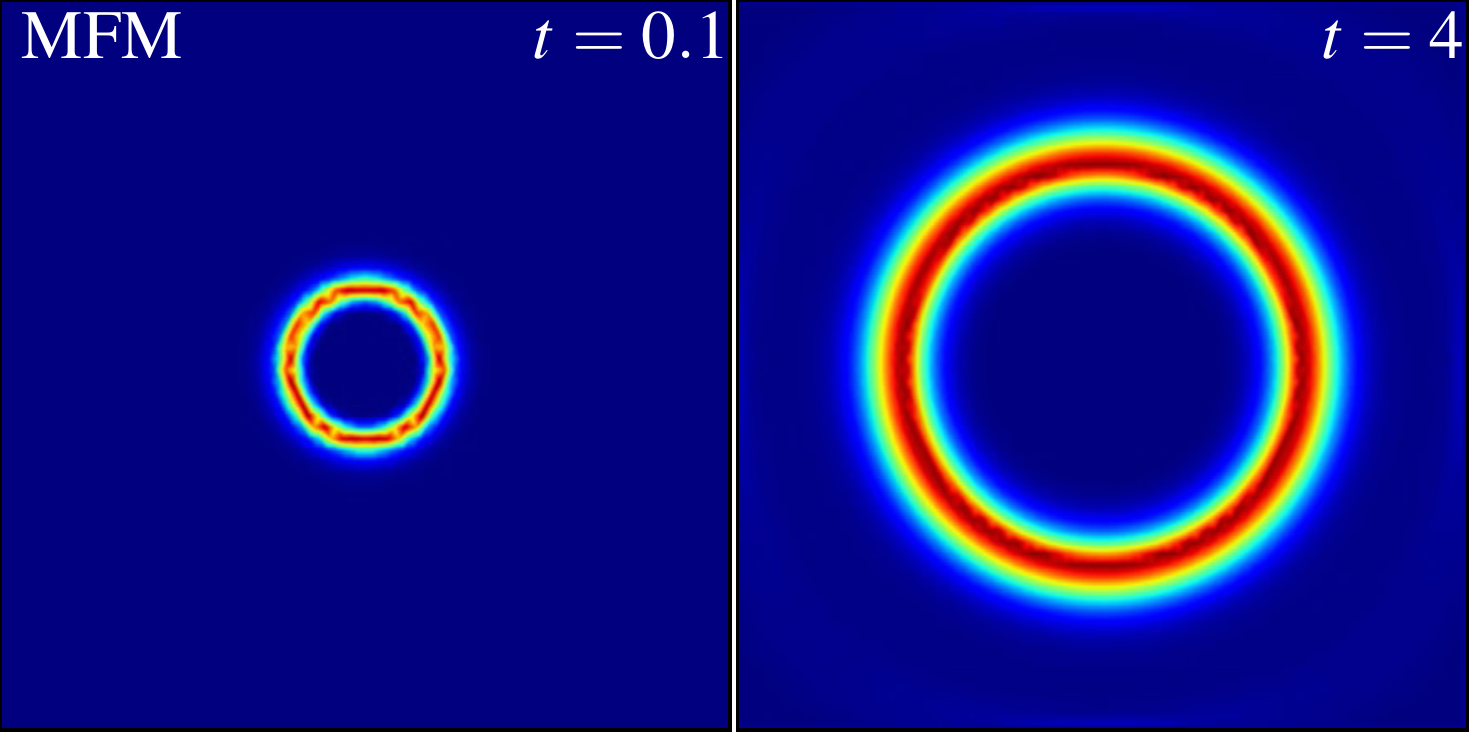}{1.0}
    \vspace{-0.5cm} 
    \caption{Diffusing ring test for radiative diffusion (\S~\ref{sec:ring.tensor}; style as Fig.~\ref{fig:pulse}), where the diffused quantity ${\bf q} = q\,\mathbb{D}_{\nu}$ is an anisotropic tensor, with $\mathbb{D}_{\nu} = \hat{\phi}\otimes\hat{\phi}$ (i.e.\ a purely azimuthal Eddington tensor). We initialize a small ($r=0.1$), thin ($\delta r=0.01$), azimuthally symmetric ring (in a box of unit length); the ring should expand with speed $\approx K/r$ and remain thin. The resolution here is $64^{2}$, and we show the result at early ($t=0.1$; {\em top}) and late ($t=4$; {\em bottom}) times. MFM captures the physically correct behavior.
        \vspacerpostplot 
    \label{fig:ring.radiation}}
\end{figure}

\vspace{-0.5cm}
\subsection{Diffusing Ring}
\label{sec:ring}

A more challenging version of this problem is diffusion with azimuthal anisotropy, following \citet{parrish.2005:mti.sims,sharma.2007:anisotropic.diffusion.stability.limiters,sharma.2011:fast.semi.implicit.anisotropic.diffusion}. In a periodic box of unit size centered on the origin and cylindrical $(r,\,\phi,\,z)$ coordiates, $\rho=u=1$, ${\bf v}=0$, and purely azimuthal magnetic fields ${\bf B}=B_{0}\,\hat{\phi}$, we initialize $q(t=0) = q_{0} + q_{1}(\,\exp[-(1/2)\,[(r-r_{0})^{2}/\delta r_{0}^{2} + \phi^{2}/\delta\phi_{0}^{2}]]$. Here $q_{0}$ and $q_{1}$ are an arbitrary background and normalization (we take $q_{0}=10^{-10}$ and $q_{1}=1$), $\delta r_{0}=0.05$ and $r_{0}=0.3$ define a Gaussian ring at radius $r_{0}$ of width $\delta r$, and $\delta\phi_{0}=0.5$ is an initial Gaussian spread about $\phi=0$ in the $\hat{\phi}$ direction. We assume ${\bf K}=K\,\hat{B}\otimes\hat{B}$.

Because the fields are purely azimuthal, the quantity should diffuse in the purely azimuthal direction, ``around the ring,'' rather than in the radial direction. At early times, this has an exact solution of the same functional form: $q(t>0) = q_{0} + q_{1}(t)\,(\,\exp[-(1/2)\,[(r-r_{0})^{2}/\delta r_{0}^{2} + \phi^{2}/\delta\phi^{2}]]$ where $\delta\phi^{2}= \delta\phi^{2}(r,\,t) = \delta\phi_{0}^{2} + 2\,K\,r^{-2}\,t$ (with normalization $q_{1}(r,\,t)=q_{1}(t=0)\,(\delta\phi_{0}/\delta\phi)$). This assumes we can neglect the periodic boundary conditions around the ring -- i.e.\ is valid for $\delta\phi\ll\pi$ (hence early times). At late times, the diffusion from both directions self-intersects on the opposite side of the ring, and there is no simple exact solution. Eventually, though, as $t\rightarrow \infty$, the system becomes isothermal within each azimuthal annulus. This is a challenging problem even in high-order grid codes \citep[see][]{parrish.2005:mti.sims}.

Fig.~\ref{fig:ring} compares the results at early and late times, at two different resolution levels. As before, MFM is able to capture the azimuthal anisotropy. Even at low resolution ($64^{2}$), there are only weak grid artifacts, but these and the amount of perpendicular diffusion improve at higher resolution. At late times, for comparison, in the fixed-grid code {\small ATHENA} on a Cartesian mesh (where the preferred direction of the grid is not the azimuthal direction), it requires going to $\sim 256^{2}$ resolution before the diffusion properly ``wraps'' into a ring at all \citep[see e.g.][]{sharma.2011:fast.semi.implicit.anisotropic.diffusion}; our $128^{2}$ case resembles a $\sim 512^{2}$ case with {\small ATHENA}. And we note that we have not aligned the particles with the anisotropy (the particles are in a regular triangular lattice).

Note that we have tested both the 2D ring version of this problem and the 3D version, where the ring becomes a cylinder elongated in the $\hat{z}$ direction and the box is periodic in that direction. The results are very similar in both cases.

\vspace{-0.5cm}
\subsubsection{Convergence}

Note that our diffusing ring setup is slightly different from that in \citet{sharma.2007:anisotropic.diffusion.stability.limiters,sharma.2011:fast.semi.implicit.anisotropic.diffusion} and \citet{kannan.2015:anisotropic.conduction.arepo}, who used this problem to measure the convergence properties of their method. We have therefore also compared initial conditions which match their choice: we initialize a step-function discontinuity with $q=q_{0}$ and $q=q_{1}$ inside or outside (respectively) of an annulus $0.25<r<0.35$ and $-\pi/6<\phi<\pi/6$. The qualitative results with all methods are identical to those shown in Fig.~\ref{fig:ring}. 

Fig.~\ref{fig:ring.convergence} quantifies the convergence (in 2D tests using this initial condition) by measuring the L1 norm of $q$ at time $t=0.2$ relative to a high-resolution solution ($2048^{2}$) interpolated to the particle positions. We find a convergence rate $L1 \propto N^{-0.9}$, close to the ideal $\propto N^{-1}$. This is competitive with and in some cases superior to the implementations studied in fixed-grid and moving-mesh codes in \citet{sharma.2007:anisotropic.diffusion.stability.limiters,sharma.2011:fast.semi.implicit.anisotropic.diffusion,kannan.2015:anisotropic.conduction.arepo}. However we caution that we are comparing to our own high-resolution solution, not an exact solution (because none exists); so systematic errors which may converge more slowly do not appear.


\vspace{-0.5cm}
\subsubsection{Tensor Diffusion: The Radiative Diffusion Case}
\label{sec:ring.tensor}

Interestingly, if we consider the radiative diffusion version of this problem, the behavior is qualitative different. Take $\mathbb{D}_{\nu}=\hat{n}\otimes\hat{n}$ and $\hat{n}=\hat{\phi}$ (define $K\equiv \lambda\,c/(\kappa_{\nu}\,\rho)$); because the anisotropy is {\em inside} the first gradient in the diffusion equation (i.e.\ we have ${\bf F} = K\,\nabla \cdot [(\hat{\phi}\otimes\hat{\phi})\,|{\bf q}|]$ as opposed to ${\bf F} = K\,(\hat{\phi}\otimes\hat{\phi}) \cdot \nabla |{\bf q}|$), the solution is distinct. For an azimuthally symmetric ${\bf U}=n_{\nu}=n_{\nu}(r)$ and $\hat{n} = \hat{\phi}$, the diffusion equation for scalar $q$ reduces to $\partial q/\partial t = -(K/r)\,\partial q/\partial r$, i.e.\ the ring expands radially with a speed $=K/r$. Fig.~\ref{fig:ring.radiation} shows the results of this test. In our MFM/MFV methods, the ring expands as expected, and the correct qualitative behavior is captured even at extremely low resolution ($\sim32^{2}$).

\begin{figure*}
\plotsidesize{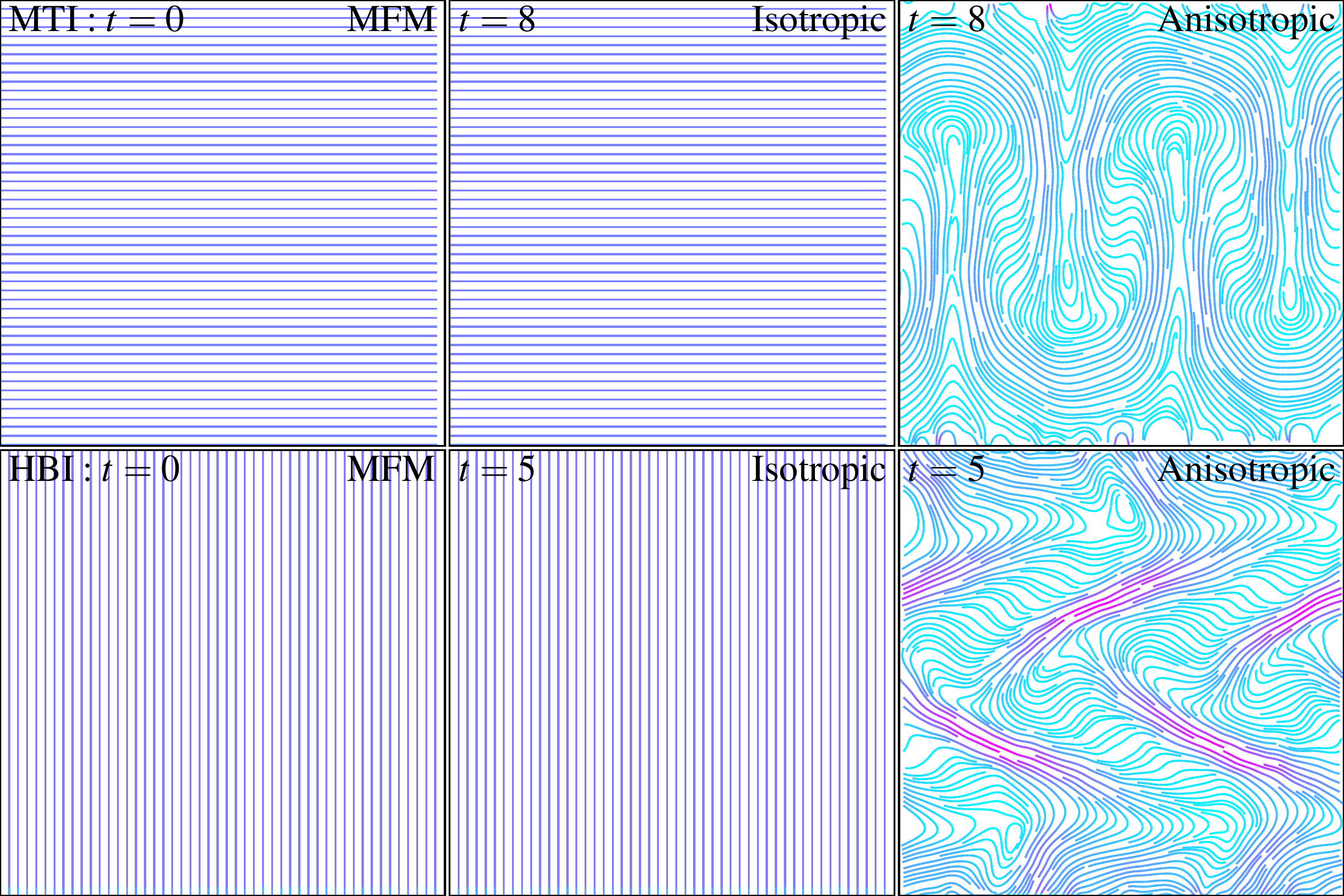}{1}
    \vspace{-0.5cm} 
    \caption{Magneto-thermal instability (MTI) and heat-flux driven bouyancy instability (HBI) tests (\S~\ref{sec:mti}), with our MFM method. 
    {\em Top:} MTI: The initial condition ({\em left}) is a 2D box of length $L=0.1$, with $256^{2}$ resolution elements, with a vertically stratified atmosphere with $dT/dz < 0$ (temperature decreases upwards), in equilibrium with a constant vertical gravitational field ${\bf g}=-\hat{z}$ and conductivity $\kappa=0.01$. Magnetic field lines are shown at different times $t$; these are initialized with trace values aligned along $\hat{x}$. Color denotes $|{\bf B}|$, increasing linearly from the minimum to maximum values at each time (cyan to purple to magenta, respectively; see Fig.~\ref{fig:hbi.evol} for quantitative values). 
    With no diffusion, or isotropic diffusion (${\bf K}=\kappa\,{\bf I}$; {\em center}), the system is stable and the field configuration should be preserved. With anisotropic diffusion (${\bf K}=\kappa\,\hat{B}\otimes\hat{B}$; {\em right}), the system is unstable and develops convection. This re-orients the field to be initially near vertical (at e.g.\ the time shown), then breaks up into turbulence and the field becomes isotropic. The characteristic timescale is the bouyancy time $\sim 1.7$ in these units.
    {\em Bottom:} HBI. The resolution and conductivity are the same but the initial atmosphere now has $dT/dz > 0$, and initial $\hat{B}=\hat{z}$. Again with no diffusion or isotropic diffusion, the system is stable. With anisotropic diffusion, the HBI amplifies transverse motions that re-orient the field to be perpendicular ($\hat{B}\rightarrow \hat{x}$).
    MFM recovers all of the expected behaviors for both instabilities with anisotropic diffusion, and correctly suppresses them with isotropic diffusion.
            \vspacerpostplot 
    \label{fig:mti.hbi}}
\end{figure*}

\begin{figure}
\plotonesize{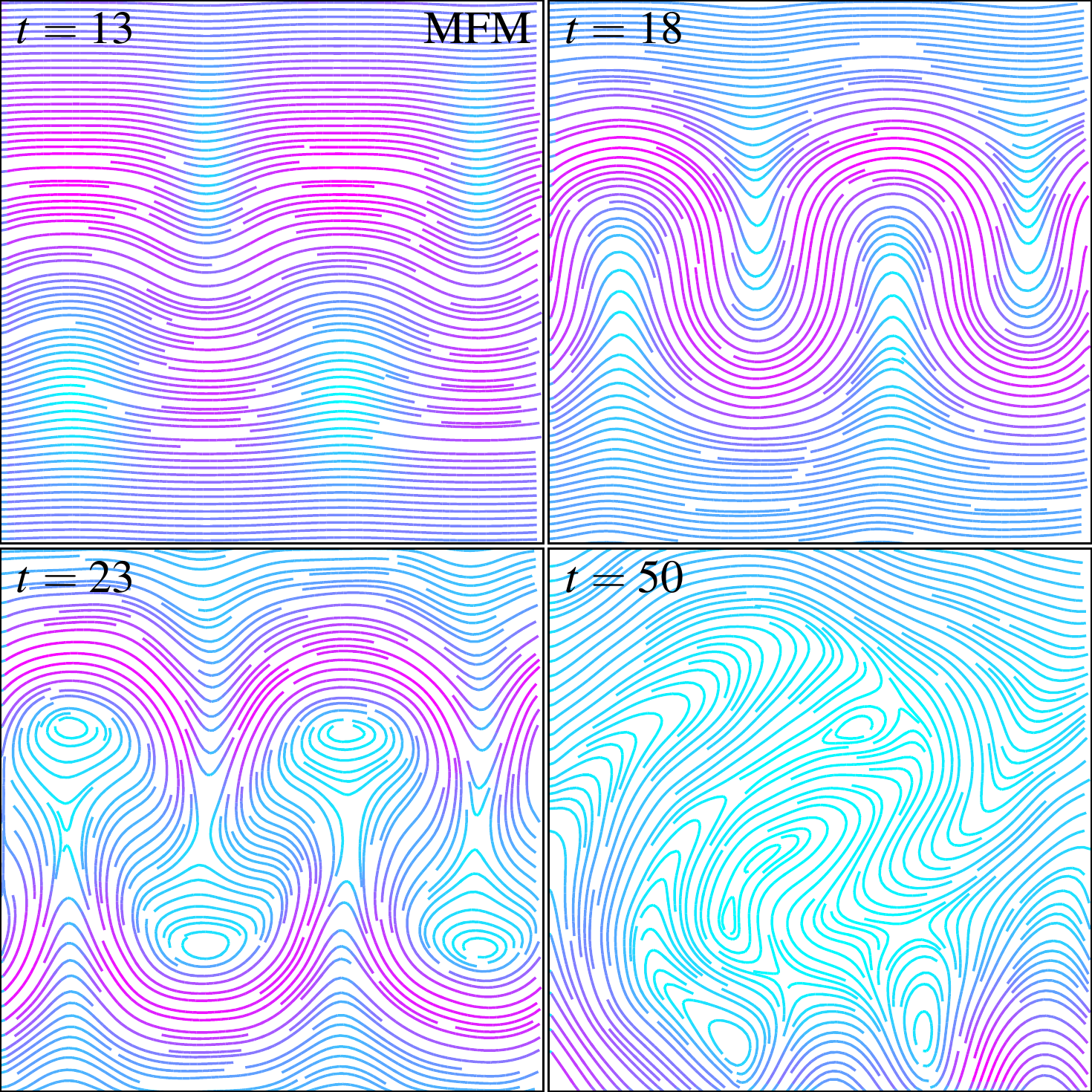}{1}
    \vspace{-0.5cm} 
    \caption{Development of the MTI in an MFM simulation at low ($32^{2}$) resolution, now with a box of side-length $L=1$. Field lines are shown as Fig.~\ref{fig:mti.hbi}. Initially small seed velocity perturbations grow in the vertical direction and become the convective cells. Once the cells reach the box boundaries ($t\gtrsim 25$, the convection is sustained by the constant-temperature, reflecting boundary conditions and produces sustained turbulence that isotropizes the magnetic field. Again, MFM can capture all the important behaviors even at very low resolution, especially for the larger box studied here which produces larger Mach numbers (here $\sim 0.03$) compared to Fig.~\ref{fig:mti.hbi}. The bouyancy time for this setup is $\sim 1.7$.
        \vspacerpostplot 
    \label{fig:mti.time}}
\end{figure}

\vspace{-0.5cm}
\subsection{Anisotropic Diffusion-Driven Instabilities: MTI \&\ HBI}
\label{sec:mti}

\begin{figure}
\plotonesize{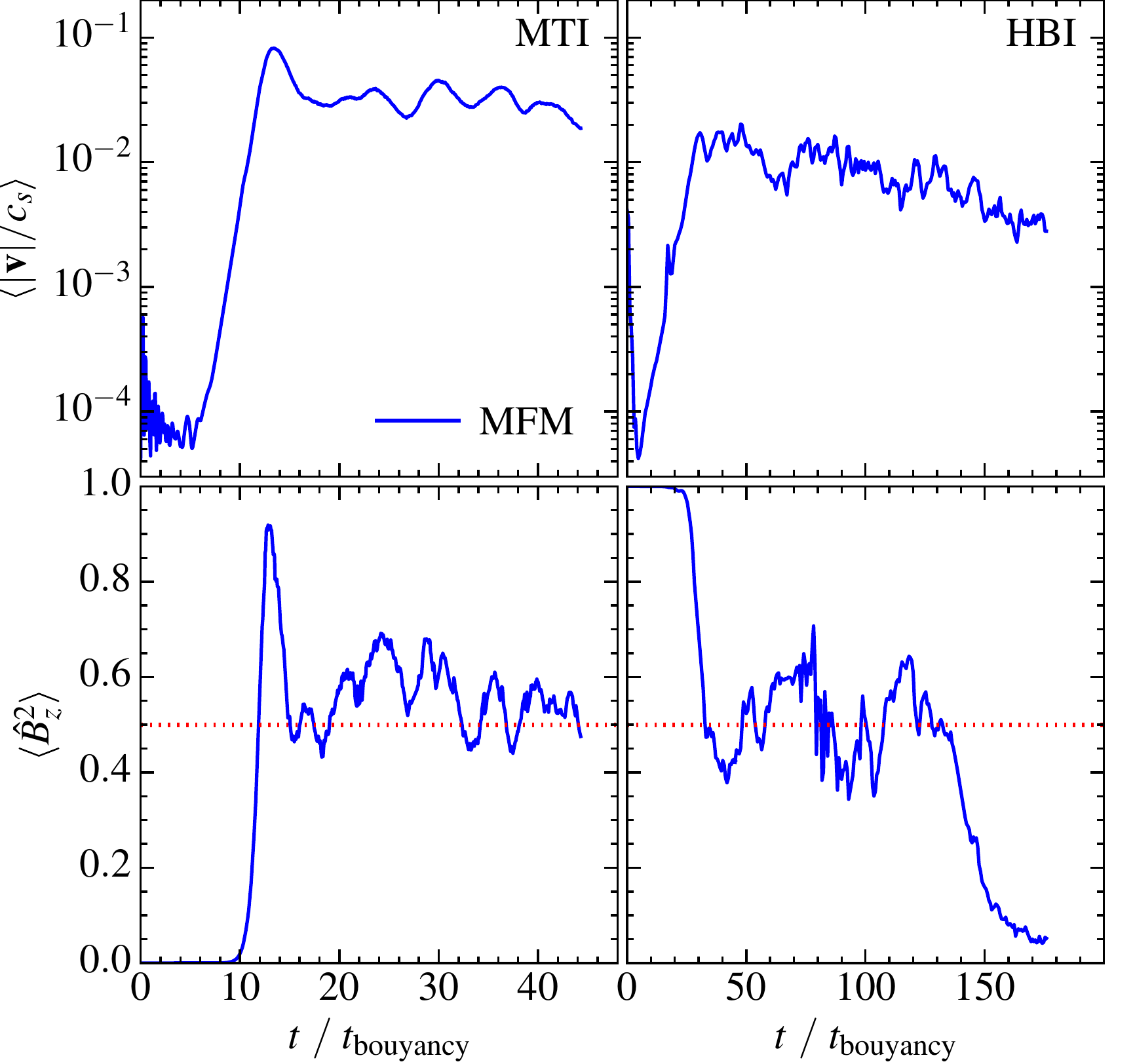}{1}
    \vspace{-0.5cm} 
    \caption{Time evolution of our low-resolution MTI ({\em left}) and HBI ({\em right}) simulations from Fig.~\ref{fig:mti.time} (\S~\ref{sec:mti.lowres}). 
    {\em Top:} rms Mach number (volume-averaged) in the center of the computational domain, as a function of time. 
    {\em Bottom:} Mean squared $z$-component of the magnetic field orientation, $\hat{B}_{z}^{2} = (\hat{B}\cdot\hat{z})^{2}$. 
    In MFM, the instabilities develops and velocities increase from their seed values rapidly around $t\sim 10-20\,t_{\rm bouyancy}$. The magnetic fields are rapidly re-oriented at the same time. In the MTI, fields go from horizontal to vertical, then the sustained convection with steady-state Mach numbers $\sim0.03-0.1$ (given the large box-size $L=1$) isotropizes the fields ({\em red dotted line} represents isotropic fields). The fluctuations about isotropy owe to the low resolution. 
    In the HBI, fields go from vertical to isotropic, but the large Mach numbers ($\sim 0.02$), low resolution, and boundary conditions here cause an overshoot that sustains fluctuations around isotropy until the velocities decay to $\lesssim 0.01\,c_{s}$, at which point the instability rapidly completes the horizontal re-orientation of the field. 
         \vspacerpostplot 
    \label{fig:hbi.evol}}
\end{figure}

We now consider two instabilities specific to anisotropically conducting plasmas: the magneto-thermal instability (MTI) and heat-flux driven bouyancy instability (HBI) \citep{balbus.2000:mti,quataert.2008:hbi}. These have been studied in various astrophysical contexts as drivers of turbulence, convection, and mechanisms to enhance or suppress conduction; our specific problem setup is motivated by the studies in \citet{parrish.2005:mti.sims,parrish.2008:mti.sims,parrish.2008:hbi.sims,mccourt.2011:bouyancy.instability.saturation,kannan.2015:anisotropic.conduction.arepo}. 

Here, we are not interested in the physics of the instabilities themselves, but they are useful numerical tests for several reasons. (1) They require accurate coupling of the anisotropic conduction to the magneto-hydrodynamics of the flow (not guaranteed in operator-split methods). (2) They are specific to anisotropic conduction and are suppressed with isotropic conduction, so directly test whether isotropic numerical diffusion can overwhelm physical diffusion. (3) They test the ability of the anisotropic conduction operator to recover small-amplitude seed perturbations. (4) They lead to non-linear, sub-sonic turbulence, which is particularly challenging for mesh-free methods (historically, SPH) to treat accurately \citep[see e.g.][]{bauer:2011.sph.vs.arepo.shocks,price:2011.sph.turb.response}, and require that our operator preserves anisotropy even in a turbulent flow.

\vspace{-0.5cm}
\subsubsection{High-Resolution Tests}
\label{sec:mti.hires}

Following \citet{parrish.2005:mti.sims,parrish.2008:mti.sims,parrish.2008:hbi.sims,mccourt.2011:bouyancy.instability.saturation,kannan.2015:anisotropic.conduction.arepo}, we initialize a 2D box in $x-z$ coordinates with an analytic constant gravitational acceleration ${\bf g} = -\hat{z}$, size $L=1/10$, resolution $256^{2}$, polytropic $\gamma=5/3$, and conductivity $K=0.01$ (${\bf K}=K\,{\bf I}$ for the isotropic case, ${\bf K}=K\,\hat{B}\otimes\hat{B}$ for the anisotropic case). For the MTI, we initialize $u=(3/2)\,(1-z/H)$ with $H=3$, and ${\bf B}=10^{-11}\,\hat{x}$, with a seed velocity perturbation\footnote{In our high-resolution tests, we use a seed velocity perturbation with magnitude $\mathcal{M}_{0}=10^{-2}$ instead of $\mathcal{M}_{0}=10^{-4}$ as in \citet{parrish.2005:mti.sims,parrish.2008:mti.sims,parrish.2008:hbi.sims}. This was chosen for computational convenience. Because we use an explicit integration method, the timestep for high resolution in a small spatial domain becomes very short ($\sim 10^{-5}$). Using the smaller seed velocity requires approximately an order-of-magnitude longer runtimes to develop non-linear behavior, so we adopt the larger seed for high-resolution tests. However we explicitly show in \S~\ref{sec:mti.lowres} below that our MFM method can accurately capture very small seed velocities even at much lower resolution $\sim 32^{2}$.} ${\bf v} = \mathcal{M}_{0}\,c_{s}\,\sin{(4\pi\,x/L)}\,\hat{z}$ with $\mathcal{M}_{0}=10^{-2}$. For the HBI, $u=(3/2)\,(1+z/H)$ with $H=2$, and ${\bf B}=10^{-11}\,\hat{z}$ and ${\bf v} = \mathcal{M}_{0}\,c_{s}\,[\sin{(3\pi\,z/L)}\,\hat{x} + \sin{(4\pi\,x/L)}\,\hat{z} ]$. The density $\rho(z)$ is initialized so that the initial pressure gradient balances gravity with $\langle\rho\rangle=1$ in the box. The domain is divided vertically into three equal sub-domains of length $L/3$; the top and bottom layers have isotropic conduction (i.e.\ are buoyantly neutral) while the central layer has anisotropic conduction (this follows the previous studies and reduces the sensitivity to boundary conditions). The boundaries are periodic in $\hat{x}$; and constant-temperature reflecting boundaries in $\hat{z}$. Note that a sharp, reflecting and conducting ``wall'' is particularly challenging to implement in mesh-free methods. We treat the reflecting boundaries as follows: a layer (3 particles deep) of boundary particles with fixed positions and temperatures is placed at $z<0$ and $z>L$. For every interaction between a boundary particles and normal particle (gradient calculation, the hydrodynamic operations, etc), the boundary particle is assumed to have all specific properties matched to the normal particle, except the temperature (fixed to the IC value), velocity and magnetic field (which follow the usual reflection rules) and density (adapted to give equal pressure, given the different temperatures).

Both of these ICs represent a stably stratified atmosphere. With no conduction, or isotropic conduction, the system should remain in equilibrium and the seed perturbations should damp. With anisotropic conduction, provided a large enough diffusivity (as chosen here) such that the system is approximately isothermal along field lines, the instabilities should grow and eventually re-orient the field lines. In the MTI (vertical temperature profile $dT/dz < 0$), the small vertical velocity grows into large, non-linear convective cells which carry the field lines and re-orient the field in the vertical direction, until the cells cross the domain and the fixed-temperature boundary conditions produce sustained turbulence. In the HBI ($dT/dz > 0$), the initial mixed perturbation generates growing separation/compression of field lines, which leads to horizontal motions that try to stretch the field lines horizontally, until the instability saturates when the field lines are horizontal and suppress further convection. In both the HBI and MTI, the characteristic timescale is the bouyancy time $|g\,\partial \ln{T}/\partial z|^{-1/2}$ ($\sim (1.7,\,1.4)$ in code units for our MTI and HBI setups, respectively). 

Fig.~\ref{fig:mti.hbi} shows the results of these tests at a few bouyancy times, using our MFM method. Recall the relatively large initial perturbation here means non-linear behavior can develop in just a couple bouyancy times, and we see that occur in the anisotropic case. The qualitative behavior of the non-linear HBI and MTI compares well to that seen in \citet{parrish.2005:mti.sims,parrish.2008:hbi.sims,mccourt.2011:bouyancy.instability.saturation}. In both MTI and HBI, the cases with isotropic diffusion show (correctly) no evidence of instability, and the initial perturbations are eventually fully damped. 


\vspace{-0.5cm}
\subsubsection{Low-Resolution Tests}
\label{sec:mti.lowres}

To explore late-time evolution and resolution effects, we now consider an initial condition which is identical to our high-resolution tests, but with lower resolution $32^{2}$ and larger box size $L=1$ (these both allow  larger timesteps), smaller seed velocity $\mathcal{M}_{0}=10^{-4}$, and larger conductivity $K=0.1$ (to preserve the desired limit where the conduction is sufficiently fast). 

Fig.~\ref{fig:mti.time} shows the time evolution of the MTI in one of these runs; we obtain similar (good) behavior in our low-resolution MTI runs. In both cases, even at $32^{2}$ resolution, the instabilities are (correctly) completely suppressed with isotropic diffusion. Quantitatively, the results from these runs are shown in Fig.~\ref{fig:hbi.evol}. 

In MFM, we see the MTI grow (despite the very low resolution), going non-linear after $\sim10-20$ bouyancy times (similar to high-resolution cases with the same seed perturbation amplitude; \citealt{parrish.2005:mti.sims,mccourt.2011:bouyancy.instability.saturation,kannan.2015:anisotropic.conduction.arepo}), at which point the convective plumes re-order the magnetic field from horizontal to vertical. At late times, the plumes break up into sustained, non-linear convection, which isotropizes the field (although this is known to be sensitive to the boundary conditions). The fluctuations around isotropy owe to the ongoing turbulence (and are smaller at higher resolution). The saturated Mach numbers $\sim 0.01-1$ are larger than those in the smaller $L=0.1$ box, as expected based on the scaling seen in \citealt{mccourt.2011:bouyancy.instability.saturation}. Both qualitative and quantitative evolution compare favorably to higher-resolution studies in fixed-grid and moving-mesh schemes \citep{mccourt.2011:bouyancy.instability.saturation,kannan.2015:anisotropic.conduction.arepo}. 

For the HBI,  the instability goes non-linear on a similar timescale to the MTI, but the Mach numbers slowly decay as the field re-aligns. Interestingly, for the $L=1$ box, there is an intermediate period of isotropic fields, before the instability completes their horizontal re-alignment; this does not appear in the $L=0.1$ box, so we suspect it owes to the larger turbulent ``overshoot'' through the unstable zone induces by the first phase of the instability, which must be damped before the re-alignment can complete. 



\begin{figure}
\plotonesize{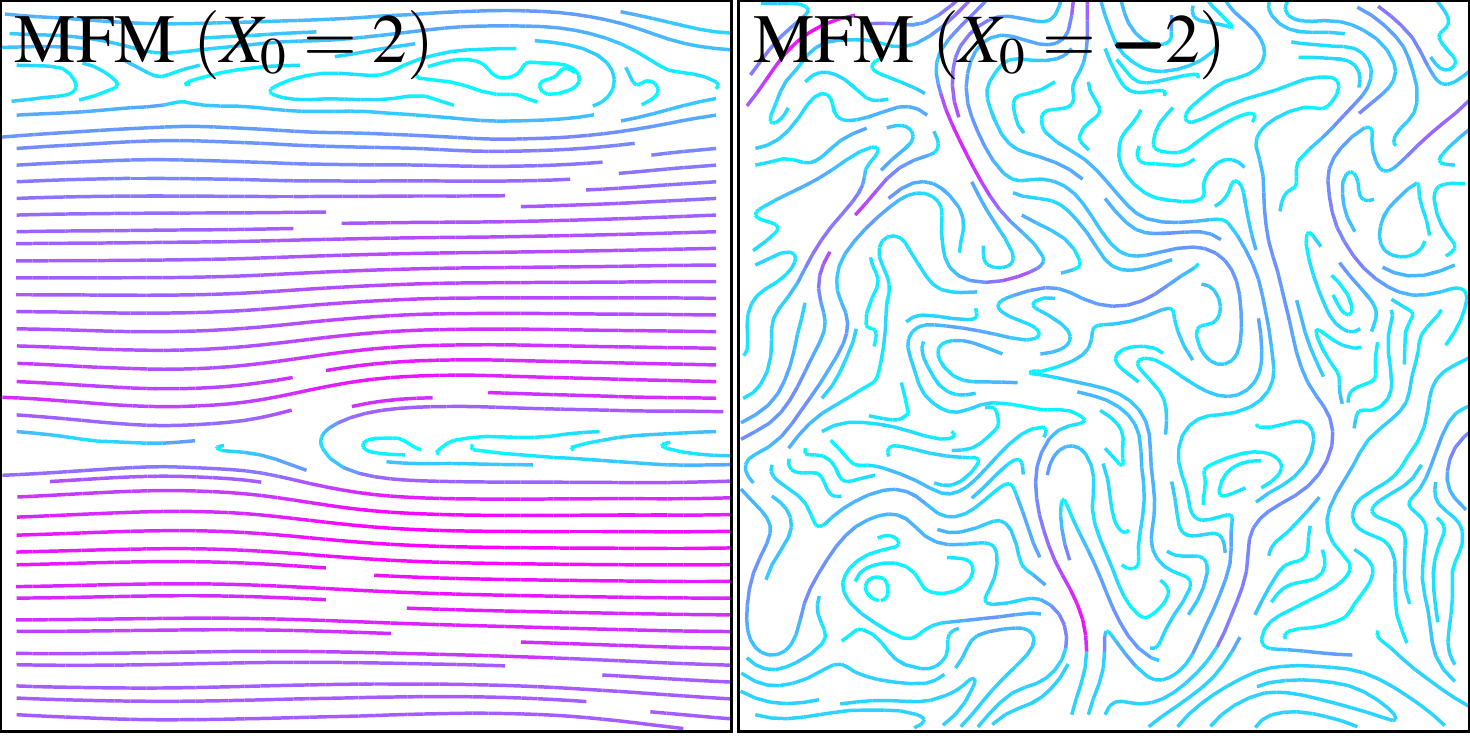}{1}
    \vspace{-0.25cm} 
    \caption{Hall MRI tests (\S~\ref{sec:hall.mri}). We compare the field lines (as Fig.~\ref{fig:mti.hbi}) well into non-linear evolution. The initial condition is a 2D $R-z$ shearing box of length $L=H$ (the pressure scale length) with $128^{2}$ resolution, a constant field $B_{0}\,\hat{z}$, trace pressure/velocity perturbations, and explicit Ohmic resistivity (with magnetic Reynolds number ${\rm Re}_{M,\,0}=10 \propto 1/\eta_{O}$) and Hall effects (with Hall parameter $X_{0} \propto 1/\eta_{H}$). Opposite signs of $X_{0}$ correspond to the identical setups with opposite signs of ${\bf B}$; this has no effects unless the Hall term is present.
    Results are shown at $t\approx6\,t_{\rm orbital}$ ($t_{\rm orbital}=2\pi/\Omega$).
    {\em Left:} For $X_{0}\ge 0$, modes should grow quickly. In the non-linear state, the fastest-growing modes become bigger than the box, leading to the horizontal channel modes seen; these increase $|{\bf B}|$ without limit.
    {\em Right:} For $-4 \lesssim X_{0} \lesssim -2$, growth is slower and smaller modes grow fastest, leading to near-isotropic MRI turbulence in the saturated state as opposed to channel modes.
            \vspacerpostplot 
    \label{fig:hall.mri.fieldlines}}
\end{figure}

\begin{figure}
\plotonesize{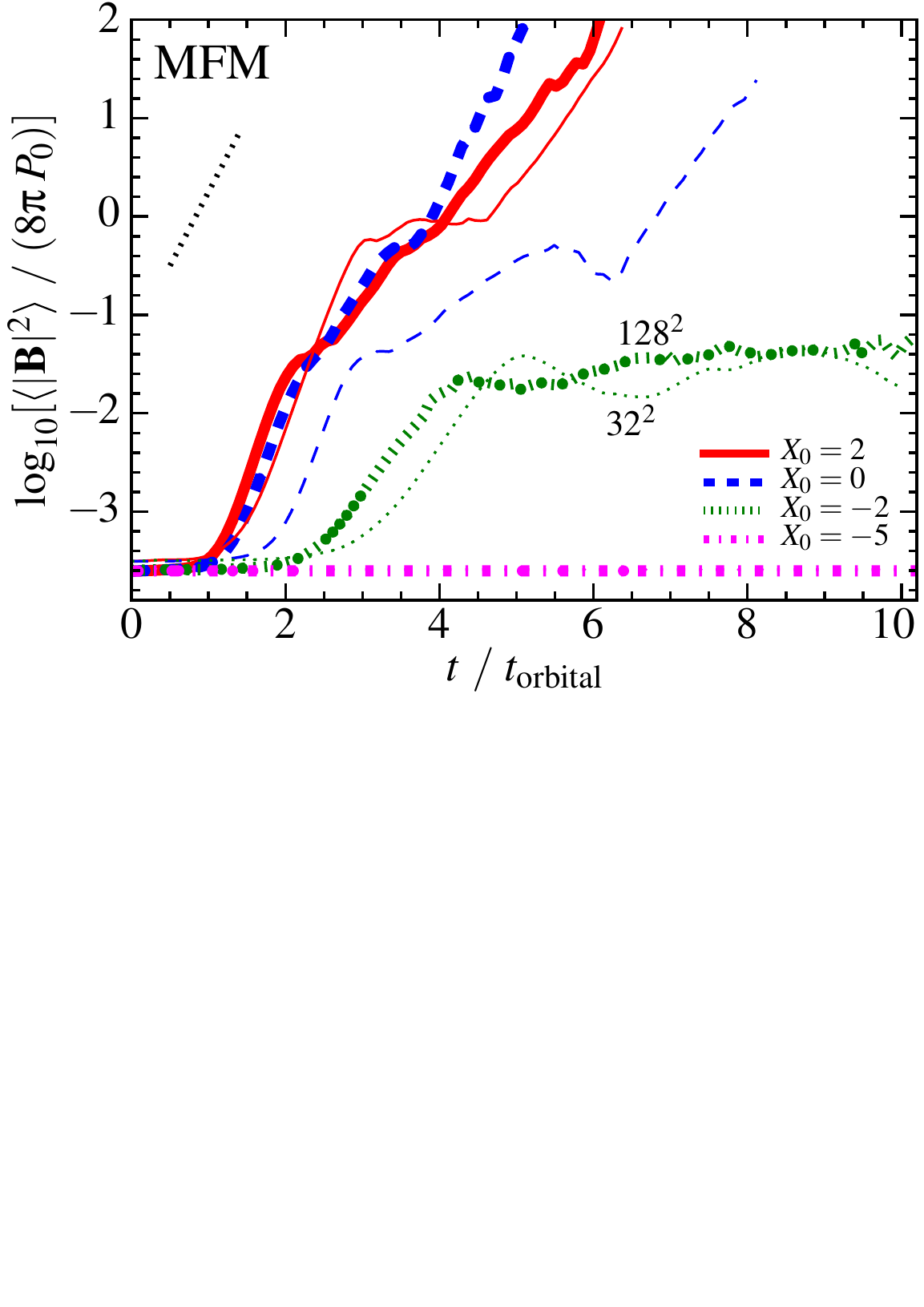}{0.95}
    \vspace{-0.25cm} 
    \caption{Growth of the volume-averaged magnetic pressure (relative to the initial thermal pressure $P_{0}$) in the mean-field Hall MRI tests. We show results for different initial Hall parameters $X_{0}=+2,\,0,\,-2,\,-5$ (as labeled), at low resolution ($32^{2}$; thin lines) and intermediate resolution ($128^{2}$; thick lines). 
    We also show the expected analytic maximum linear growth rate ($|{\bf B}|^{2} \propto \exp(1.5\,t\,\Omega)$; dotted black line), which should be close to the simulated growth rate for $X_{0}\ge0$. 
    The growth for $X_{0}\ge 0$ is similar to the ideal MHD case (as expected), with growth rates in good agreement with the analytic expectation for $128^{2}$ resolution, and late-time formation of channel modes (Fig.~\ref{fig:hall.mri.fieldlines}). Growth rates are suppressed at very low resolution, as expected from the MRI studies in \papertwo. For $X_{0}=-2$, linear growth is suppressed and the magnetic field strength saturates when the system saturates in MHD turbulence (also as expected). For $X_{0}=-5$, the system should be stable against MRI growth; we confirm this and that the initial seed noise damps at the expected rate (decay rate $\sim \Omega$).
    All the results here are in good agreement with well-tested Eulerian codes (see e.g.\ \citealt{sano.2002:hall.mri.test.problems}, Fig.~5). 
            \vspacerpostplot 
    \label{fig:hall.mri.growth}}
\end{figure}

\vspace{-0.5cm}
\subsection{Hall MRI}
\label{sec:hall.mri}

As discussed in \S~\ref{sec:diffusion.sheet.special}, the Hall effect presents unique numerical challenges; therefore, we now consider the magneto-rotational instability (MRI) with a Hall term and Ohmic resistivity (``Hall MRI''). This again involves the effects of anisotropic diffusion on plasma instabilities and turbulence. The effects of the Hall term are especially important in the context of proto-stellar and proto-planetary disks \citep[for reviews, see][]{bablus:mhd.angular.momentum.transport.review,wardle.2007:mhd.protoplanetary.disk.review}, and have been studied in detail using Eulerian methods \citep{flock.2011:3d.nonideal.mhd.disk.sims,simon.2011:nonideal.mhd.stratified.disk.effects,bai:2011.nonideal.effects.on.mri.accretion.in.planetary.disks,simon.2015:mri.driven.protoplanetary.disk.accretion}. 

The MRI itself is astrophysically interesting in a wide range of contexts involving magnetized disks, and is numerically particularly interesting because it has historically proven challenging for SPH methods to correctly capture its growth \citep[see e.g.][]{rosswog:2007.sph.mhd,price:2008.sph.mhd.star.clusters,dolag:2009.mhd.gadget}. In \papertwo\ and \citet{hopkins:cg.mhd.gizmo}, we consider extensive studies of the MRI in ideal MHD ($X_{0}=0$, ${\rm Re}_{M,\,0}\rightarrow\infty$) in {\small GIZMO}. We showed that our MFM and MFV methods recover the correct linear growth rates and non-linear properties in good agreement with well-tested higher-order Eulerian codes such as {\small ATHENA}. We also showed that at least some SPH schemes were capable of doing the same (albeit with greater noise), if a larger neighbor number is used. 

Here, we follow \citet{sano.2002:hall.mri.test.problems} and consider a simplified test problem. We adopt a 2D axisymmetric shearing box from \citet{guan:2008.shearing.box.mri}; this is locally Cartesian with one coordinate ($x$) representing the radial direction and the other coordinate ($z$) the vertical direction, representing a small, azimuthally symmetric ``ring'' about the midplane of a disk in a Keplerian potential. Details of the boundary conditions and gravitational terms, as implemented in {\small GIZMO}, are in \papertwo\ (this is the same box used for the MRI simulations therein). The box has initially constant density $\rho_{0}=1$, side-length $L=1=H$ (where $H\equiv (2/\gamma)^{1/2}\,c_{s,\,0}/\Omega$ is the scale-height), orbital frequency $\Omega=1$, mean pressure $P_{0}=c_{s,\,0}^{2}/\gamma=\Omega^{2}/2$ ($\gamma=5/3$), spatially uncorrelated random pressure and velocity fluctuations with uniform distribution and $|\delta P|/P_{0}=|\delta {\bf v}|/c_{s,\,0} \le 0.5\times10^{-2}$, and uniform vertical field ${\bf B}=B_{0}\,\hat{z}$. The MRI is then characterized by three numbers, the plasma beta $\beta_{0} \equiv P_{0}/(v_{A,\,0}^{2}\,\rho_{0}) = 3200$ ($v_{A,\,0}\equiv |B_{0}|/(4\pi\,\rho_{0})^{1/2}$); the magnetic Reynolds number ${\rm Re}_{M,\,0} \equiv v_{A,\,0}^{2}/(\eta_{O}\,\Omega) = 10$, which determines the Ohmic resistivity $\eta_{O}$; and the Hall parameter $X_{0} \equiv c\,B_{0}\,\Omega/(2\pi\,e\,n_{e,\,0}\,v_{A,\,0}^{2})$, which determines $\eta_{H} = |X_{0}|\,v_{A,\,0}^{2}\,(2\,\Omega)^{-1}\,(|{\bf B}|/|B_{0}|)\,(\rho_{0}/\rho)$ (we assume the free electron fraction is constant).

Figs.~\ref{fig:hall.mri.fieldlines}-\ref{fig:hall.mri.growth} show the resulting evolution of the magnetic fields, for $X_{0}=+2,\,0,\,-2,\,-5$. For $X_{0}=0$ (no Hall term), we simply have the MRI with explicit resistivity. In 2D shearing boxes, the fastest-growing modes for ${\rm Re}_{M,\,0}=10 \gg1$ should have growth rates only slightly smaller than the $0.75\,\Omega$ expected in the ideal limit; once non-linear, they should form an inverse cascade until horizontal channel modes appear which grow without limit. For $X_{0}>0$, the behavior should be essentially identical, with slightly faster mode growth at higher $X_{0}$. When $X_{0}<-4$, the system is fully stable against the MRI, and the initial perturbations should decay. At intermediate $-4<X_{0}<0$, the MRI should grow but with a suppressed maximum growth rate (for finite ${\rm Re}_{M}$). Within this range, as $X_{0}$ becomes more negative (larger $|X_{0}|$), smaller-scale modes grow faster, until for $X_{0}\lesssim -2$ there is no critical scale at all; because of this, the system can saturate for many orbital periods with steady-state MRI turbulence (as the growing small-scale, high-$k$ modes prevent the formation of low-$k$ channel modes). We confirm all these behaviors here (compare Fig.~5 in \citealt{sano.2002:hall.mri.test.problems}). As expected and shown in detail in \papertwo, the linear growth rates are suppressed at very low resolution ($\sim 32^{2}$), but rapidly approach the analytic solution at higher resolution.  Moreover, we note that we have re-run every problem in \citet{sano.2002:hall.mri.test.problems} with our MFM method (varying $\beta_{0}$, ${\rm Re}_{M,\,0}$, $X_{0}$, and the box size; considering zero net-field cases; and their whistler wave test problem) and confirm identical qualitative behavior.

{Note that, for $-4<X_{0}<0$, given our setup, the instantaneous Hall parameter $X\propto |{\bf B}|^{-1}$ and magnetic Reynolds number ${\rm Re}_{M}\propto |{\bf B}|^{2}$, so as modes grow non-linear ($|{\bf B}|$ increases), the system moves closer to the ideal MHD limit. Because our problem has finite ${\rm Re}_{M,\,0}=10$, if we begin with $-2.1<X_{0}<0$, there is a fastest-growing mode, with relatively large wavelengths ($\lambda_{\rm max}\approx 0.03\,H,\,0.09\,H$ for $X_{0}=-2,\,-1$). During the turbulent phase, these modes can increase $|{\bf B}|$ non-linearly, which in turn increase the fastest-growing mode growth rate and wavelength, and suppress the growth of the smaller-scale modes. This can eventually trigger a runaway inverse cascade that produces the channel modes seen for $X_{0}\ge0$. We find that, if $\lambda_{\rm max}$ is well-resolved, this occurs eventually if $X_{0}\ge -2$ (for ${\rm Re}_{M,\,0}=10$), although it in some cases requires $\sim 100$ orbital times. For $X_{0}<-2.1$, we confirm the turbulence remains essentially indefinitely.}

\vspace{-0.5cm}
\subsection{Other Multi-Physics Problems}

We have also vetted our algorithms in other non-linear problems where diffusion is not necessarily the primary physics. For example, we have considered a Sedov-Taylor blastwave with thermal conduction, and a shocktube with physical viscosity following \citep{sijacki.2006:viscosity.sph.cluster.simulations}. We find good agreement between our MFM method and {\small ATHENA}; small differences are dominated by the known numerical differences in their solution of the normal MHD equations (see \paperone-\papertwo), not the sub-dominant diffusion terms. Therefore we do not consider these problems good tests of the diffusion treatment in itself (and do not study them further here). But they do serve as a validation that the diffusion operators here behave properly when coupled to additional dynamics. 

A methods paper specifically devoted to the implementation of radiation transport is in preparation, where we compare the flux-limited diffusion, optically-thin variable Eddington tensor \citep[OTVET;][]{gnedin.abel.2001:otvet}, and M1 moment closure approximations and consider several dynamical test problems designed to study the radiation-hydrodynamics of ionizing photons (Khatami et al., in prep). A similar detailed study of the cosmic ray implementation, including cosmic ray pressure effects and the role of cosmic ray streaming (as well as diffusion) is also in preparation (Chan et al., in prep). 

As a ``stress test'' of our implementations, we have also run full cosmological simulations of galaxy formation using the Feedback in Realistic Environments (FIRE) models \citep{hopkins:2013.fire,faucher-giguere:2014.fire.neutral.hydrogen.absorption,ma:2015.fire.escape.fractions,ma:2015.fire.mass.metallicity,chan:fire.dwarf.cusps,muratov:2015.fire.winds,onorbe:2015.fire.cores}, with {\small GIZMO} in its MFM mode. These simulations include gas, stars, super-massive black holes, and dark matter, self-gravity, cosmological integrations, cooling physics and gas chemistry, star formation, and feedback from stars in the forms of photo-heating, radiation pressure, stellar winds, and supernovae. We have considered cases with MHD and using the Spitzer-Braginskii coefficients for anisotropic conduction and viscosity, and Smagorinski eddy diffusion for metals. The results are presented in \citet{su:2016.weak.mhd.cond.visc.turbdiff.fx}. For our purposes here, since there is no known ``right'' answer for such simulations, we consider these only to be useful tests of numerical stability under extreme conditions. Critically, we see no evidence for numerical instability or unphysical features owing to the addition of anisotropic diffusion operators in these simulations.

\vspace{-0.5cm}
\section{Discussion}

We present numerical discretizations of general anisotropic tensor diffusion operators for Lagrangian hydrodynamics methods, specifically for both recently-developed meshless finite-mass or finite-volume (MFM/MFV) Godunov schemes. We implement these in the multi-method code {\small GIZMO}, with the specific implementations relevant for passive scalar diffusion, non-ideal MHD (Ohmic resistivity, the Hall effect, and ambipolar diffusion), sub-grid ``turbulent eddy diffusion'' models, anisotropic conduction and viscosity (shear/bulk or Braginskii), cosmic ray diffusion and streaming, and anisotropic radiation diffusion (with a variable Eddington tensor). 

We consider a variety of test problems. In all cases, our finite-element MFM/MFV schemes can produce accurate solutions, even at low resolution, and are numerically stable. The schemes are also manifestly conservative. They are able to recover correctly the anisotropic cases, up to and including complete suppression with perpendicular magnetic fields. We show this is true regardless of the local particle arrangement/disorder (even in ``worst case'' scenarios where the particle arrangement is totally random), and regardless of the ``neighbor number'' in the spline. For some cases of great astrophysical interest, e.g.\ diffusion across a moving contact discontinuity which is not aligned with the grid, these methods (by virtue of being Lagrangian and mesh-free) may exhibit substantially reduced numerical diffusion compared to non-moving, grid-based codes (e.g.\ AMR methods) at the same resolution. The MFM/MFV methods are able to capture subtle instabilities driven by anisotropic diffusion (e.g.\ the magneto-thermal and heat-flux driven bouyancy instabilities, and Hall MRI).

The particular form we adopt for the flux-limited Riemann problem is non-trivial, and should be useful for other explicit anisotropic diffusion methods. Stabilizing these methods without introducing excessive numerical diffusion is challenging, especially in irregular/unstructured meshes or mesh-free configurations. The method we propose has the advantage that it trivially generalizes to arbitrarily high-order (and complicated) gradient estimators and slope-limiters, as well as higher-order reconstruction of the gradients at the faces (we simply replace the left and right states in Eqs.~\ref{eqn:fhll.1}-\ref{eqn:fhll.2} with their appropriate values); it also admits arbitrarily complex tensors for both the diffusivity and diffused quantities (as opposed to many methods which are specific to scalar diffusion). And the flux computation is pair-wise and negligible in cost compared to the MHD Riemann problem solution. It therefore is of interest not just to mesh-free methods but also moving-mesh and AMR methods. 

At least some aspects of our new method here can also be applied to SPH. Because {\small GIZMO} is a multi-method code which includes an optional SPH solver, we present an application of this in Appendix~\ref{sec:sph}. There we note that some of the low-order errors in other SPH-based anisotropic diffusion formulations can be resolved by application of the higher-order matrix gradient estimators here, but this in turn requires application of a flux-limiter akin to what we use here, to stabilize the method. We show that the resulting ``integral-Godunov'' SPH behaves well in at least a subset of our tests. 

The major dis-advantage of our method is that it does not trivially generalize for implicit solvers (see Appendix~\ref{sec:implicit}). This is a subject that merits investigation in future work, since implicit methods can often provide a large speed boost to certain types of problems. However, the methods here are amenable to significant acceleration via super-timestepping, which can provide a comparable speedup as described in Appendix~\ref{sec:super.timestep}.

\vspace{-0.5cm}
\acknowledgments 
We thank Eliot Quataert for several helpful suggestions of interesting test problems. Support for PFH was provided by the Gordon and Betty Moore Foundation through Grant \#776 to the Caltech Moore Center for Theoretical Cosmology and Physics, an Alfred P. Sloan Research Fellowship, NASA ATP Grant NNX14AH35G, and NSF Collaborative Research Grant \#1411920. Numerical calculations were run on the Caltech compute cluster ``Zwicky'' (NSF MRI award \#PHY-0960291) and allocation TG-AST130039 granted by the Extreme Science and Engineering Discovery Environment (XSEDE) supported by the NSF. 
\\

\vspace{-0.2cm}
\bibliography{/Users/phopkins/Dropbox/Public/ms}

\begin{appendix}

\vspace{-0.5cm}
\section{Application to SPH}
\label{sec:sph}

As described in \paperone\ \&\ \papertwo, {\small GIZMO} is a multi-method code: users can run with our meshless Godunov (MFM or MFV) hydrodynamic methods detailed in the main text, or SPH, if desired.

Isotropic diffusion in SPH is generally well-handled by the hybrid-derivatives formulation of \citet{brookshaw.1985:diffusion.sph.isotropic}; we confirm this in {\small GIZMO}. Anisotropic cases are more difficult. \citet{espanol:tensor.mixed.sph.derivatives} derived a second-order accurate generalization of the \citet{brookshaw.1985:diffusion.sph.isotropic} method, but \citet{petkova.springel.2009:otvet.gadget} and \citet{arth.2014:anisotropic.conduction.sph.gadget} show that it is (catastrophically) numerically unstable when the degree of anisotropy is large. They both derived the minimum correction/dissipation term necessary to stabilize the equation, giving the hybrid-derivatives SPH equation: 
\begin{align}
\label{eqn:sph.eqn}\frac{d(V\,{\bf U})_{a}}{d t} =& \frac{m_{a}}{\rho_{a}}\sum_{b}\,\frac{m_{b}}{\rho_{b}}\,{\bf x}_{ab}^{T}\,\left[ \frac{{\bf K}_{a}+{\bf K}_{b}}{|{\bf x}|_{ab}^{2}} \right] \cdot\\
\nonumber &\left[\frac{\left(\nabla_{a}\,W_{ab} + \nabla_{b}\,W_{ab}\right)}{2} \otimes \left({\bf q}_{b} - {\bf q}_{a} \right) \right] 
\end{align}
where ${\bf x}_{ab}={\bf x}_{a}-{\bf x}_{b}$. 
Unfortunately, both \citet{petkova.springel.2009:otvet.gadget} and \citet{arth.2014:anisotropic.conduction.sph.gadget} show that this is fundamentally unable to capture significant anisotropy -- essentially, Eq.~\ref{eqn:sph.eqn} represents a linear combination of $\sim$1 part isotropic diffusion for every $\sim$5 parts anisotropic diffusion. Thus, the incorrect physical equations are being solved, and the errors (the isotropic part) do not converge with resolution. Even in a simple diffusing slab, this gives systematically incorrect solutions, and effects which depend on anisotropy (the MTI, HBI, diffusing ring) cannot be captured. 

A more accurate treatment follows the ``gradients of gradients'' method as in \citet{sijacki.2006:viscosity.sph.cluster.simulations}. Because second-kernel derivatives are unstable and catastrophically noisy, and successive application of the first-order consistent SPH gradient operator yields a non-conservative formulation \citep{morris:1996.sph.stability,price:2012.sph.review}, one applies instead a pair of conjugate first-order and zeroth-order SPH gradient estimators, giving: 
\begin{align}
\nonumber \frac{d(V\,{\bf U})_{a}^{\rm SPH}}{d t} = \sum_{b}\,{m_{a}m_{b}}
{\Bigl(}&
\frac{{\bf K}_{a}\cdot\langle \nabla \otimes {\bf q} \rangle_{a}}{\Omega_{a}\,\rho_{a}^{2}}\cdot \nabla_{a}W_{ab} \\
\label{eqn:sph.grad.grad} &+ \frac{{\bf K}_{b}\cdot\langle \nabla \otimes {\bf q} \rangle_{b}}{\Omega_{b}\,\rho_{b}^{2}}\cdot\nabla_{b}W_{ab} 
{\Bigr)}
\end{align}
where $\Omega_{a}\sim 1$ is a correction term accounting for variations in the smoothing length $h$ (see \citealt{springel:entropy}), and the SPH first-order consistent derivative operator $\langle \nabla \otimes {\bf q} \rangle_{a}$ is given by 
\begin{align}
\label{eqn:sph.first.deriv} \langle \nabla \otimes {\bf q} \rangle^{\alpha\beta...}_{a,\,{\rm SPH}} &= \sum_{c}\,\frac{m_{c}}{\Omega_{a}\rho_{a}}\,({\bf q}_{c}-{\bf q}_{a})^{\beta...}\,(\nabla_{a}W_{ac})^{\alpha}
\end{align}
Eq.~\ref{eqn:sph.grad.grad} has primarily been used for {\em isotropic} diffusion \citep{sijacki.2006:viscosity.sph.cluster.simulations,tsukamoto.2013:ohmic.dissipation.sph.super.timestepping,wurster.2014:sph.mhd.ambipolar.diffusion}, but recently was applied for anisotropic non-ideal MHD terms in \citet{tsukamoto.2015:sph.hall.mhd.protoplanet.disks,wurster.2016:non.ideal.mhd.braking}. However, two problems remain. {\bf (1)} Eq.~\ref{eqn:sph.first.deriv} has a low-order error: for a constant gradient ($q=q_{0}+q^{\prime}\,x$), the gradient returned by Eq.~\ref{eqn:sph.first.deriv} is $\nabla q = q^{\prime}\,(\hat{x} + \boldsymbol{\epsilon})$ where the error term $\boldsymbol{\epsilon} \sim \mathcal{O}(h^{0})$ depends only on the particle arrangement in the kernel
($\boldsymbol{\epsilon}_{x} = -1 + \sum x_{ca}\,f_{ca}$, $\boldsymbol{\epsilon}_{y,\,z} = \sum (y,\,z)_{ca}\,f_{ca}$ where $f_{ca}\equiv (x_{ca}/|{\bf x}_{ca}|)\,(m_{c}/\Omega_{a}\rho_{a})\,\partial W_{ca}/\partial |{\bf x}_{ca}|$) so does not converge with resolution. {\bf (2)} Similar, Eq.~\ref{eqn:sph.grad.grad} is zeroth-order, giving error terms $\sim \mathcal{O}(q^{\prime\prime}\,h^{0}) + \mathcal{O}(q^{\prime}\,h^{-1})$. 

\begin{figure}
 \begin{tabular}{c}
  \includegraphics[width=0.9\columnwidth]{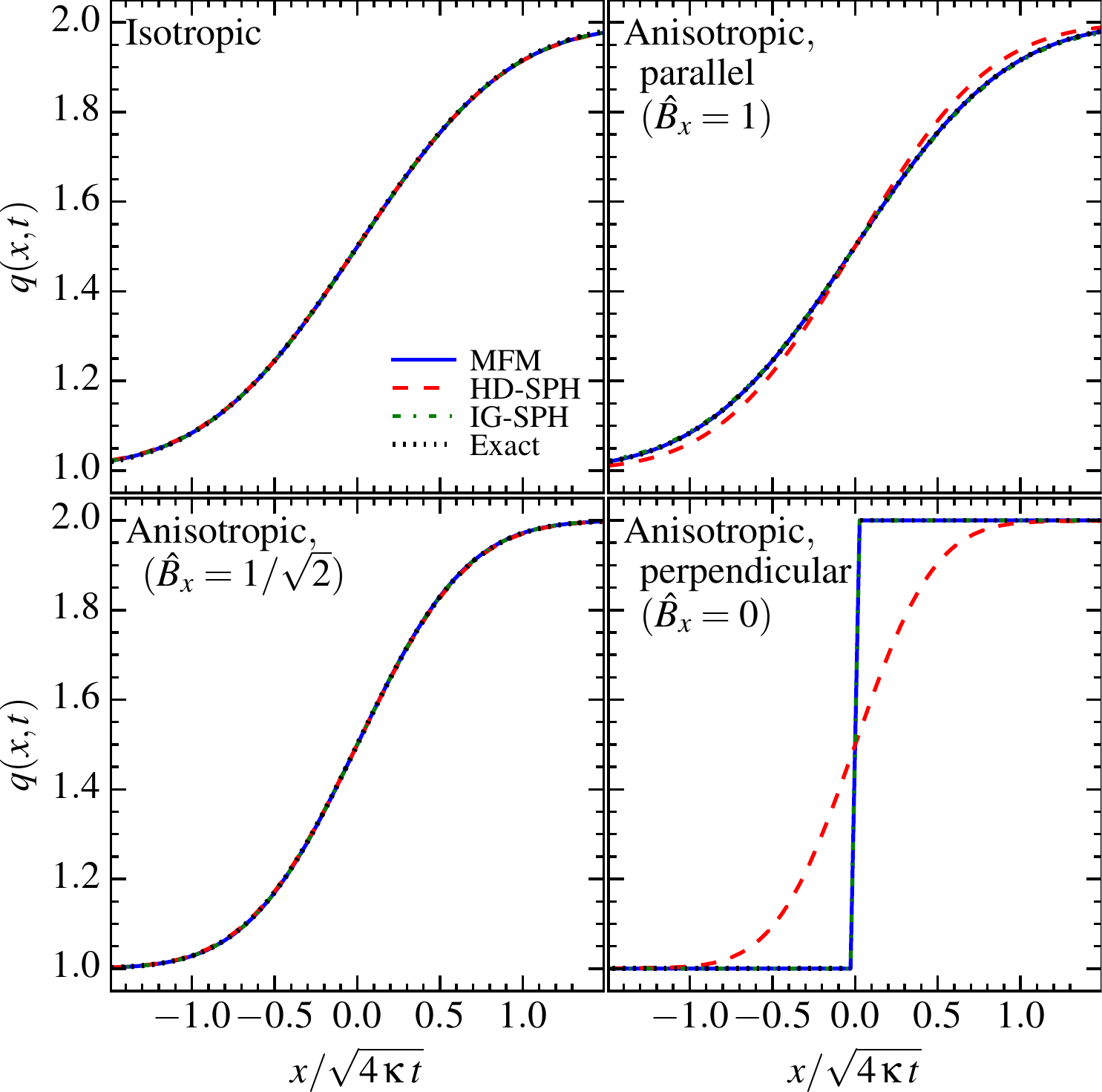} \\
  \includegraphics[width=0.9\columnwidth]{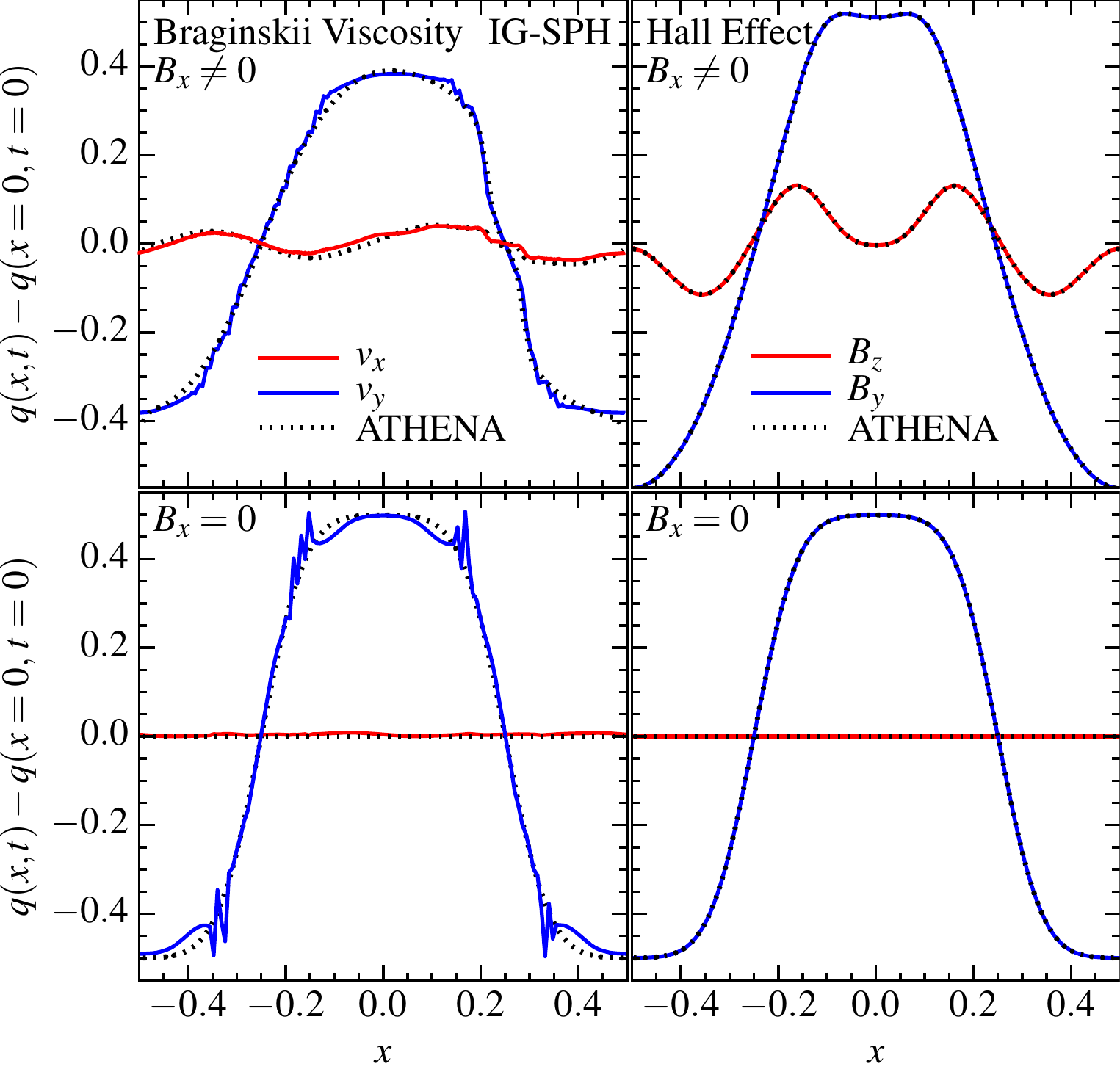} 
 \end{tabular}
    \vspace{-0.25cm}
    \caption{{\em Top:} Diffusing sheet test as Fig.~\ref{fig:diff.sheet.basic}, but comparing smoothed-particle hydrodynamics (SPH) methods as described in \S~\ref{sec:sph}. The ``hybrid derivatives'' HD-SPH method (Eq.~\ref{eqn:sph.eqn}) is unable to capture proper anisotropy, as is well-known, so we will not consider it further. Our new ``integral-Godunov'' SPH method, which borrows the higher-order matrix gradient estimators and same flux-limited Riemann solution from the MFM method in the main text, gives very similar results to MFM. {\em Bottom:} Braginskii viscosity and Hall effect as Fig.~\ref{fig:diff.brag}, in IG-SPH. The hall effect is captured well; there is some noise in Braginskii viscosity owing to the residual ``E0'' SPH errors and interaction between the explicit and ``artificial'' viscosity in SPH. 
        \vspacerpostplot 
    \label{fig:sph.sheets}}
\end{figure}

\begin{figure}
 \begin{tabular}{c}
  \includegraphics[width=0.9\columnwidth]{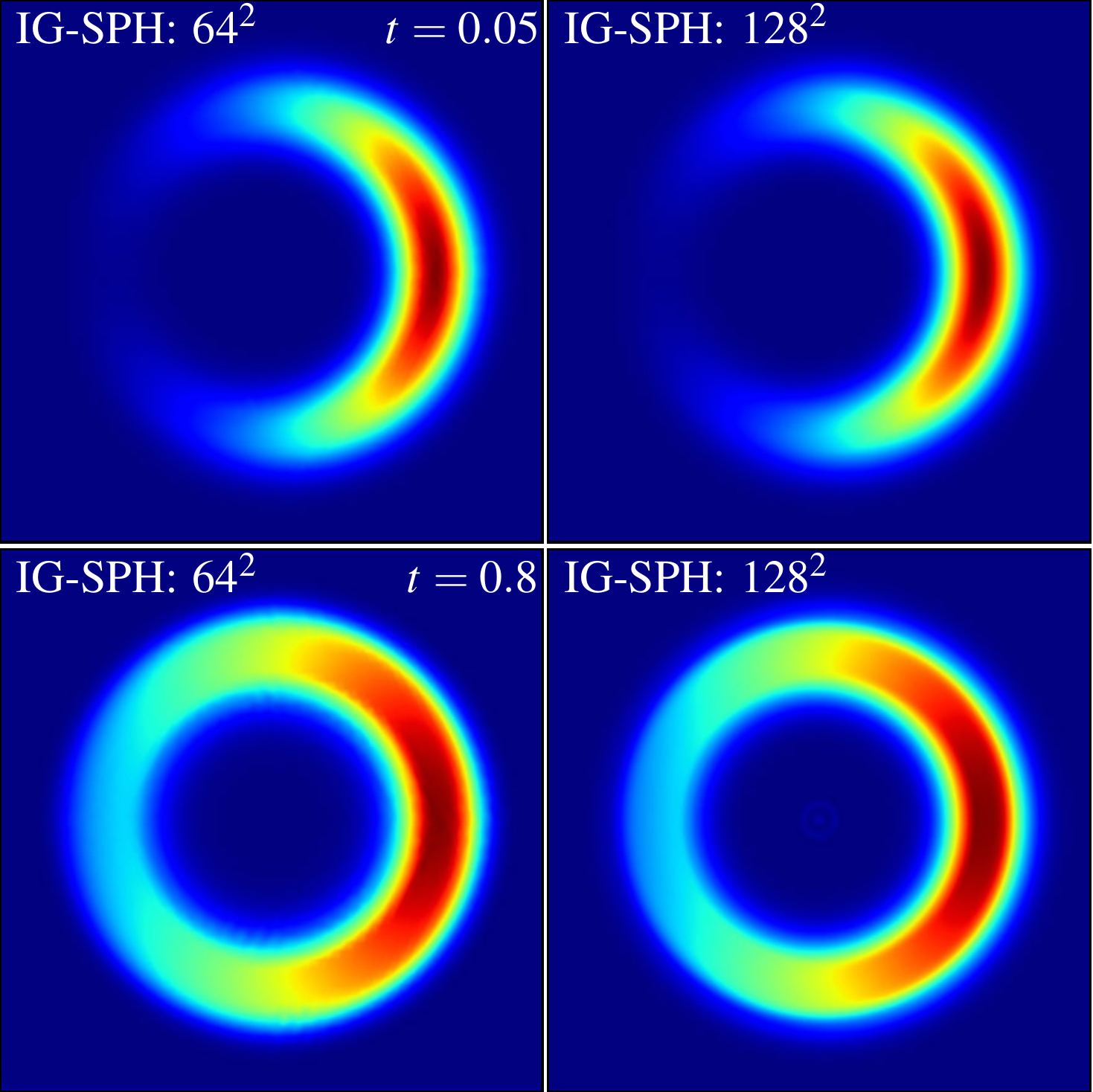} \\
  \includegraphics[width=0.9\columnwidth]{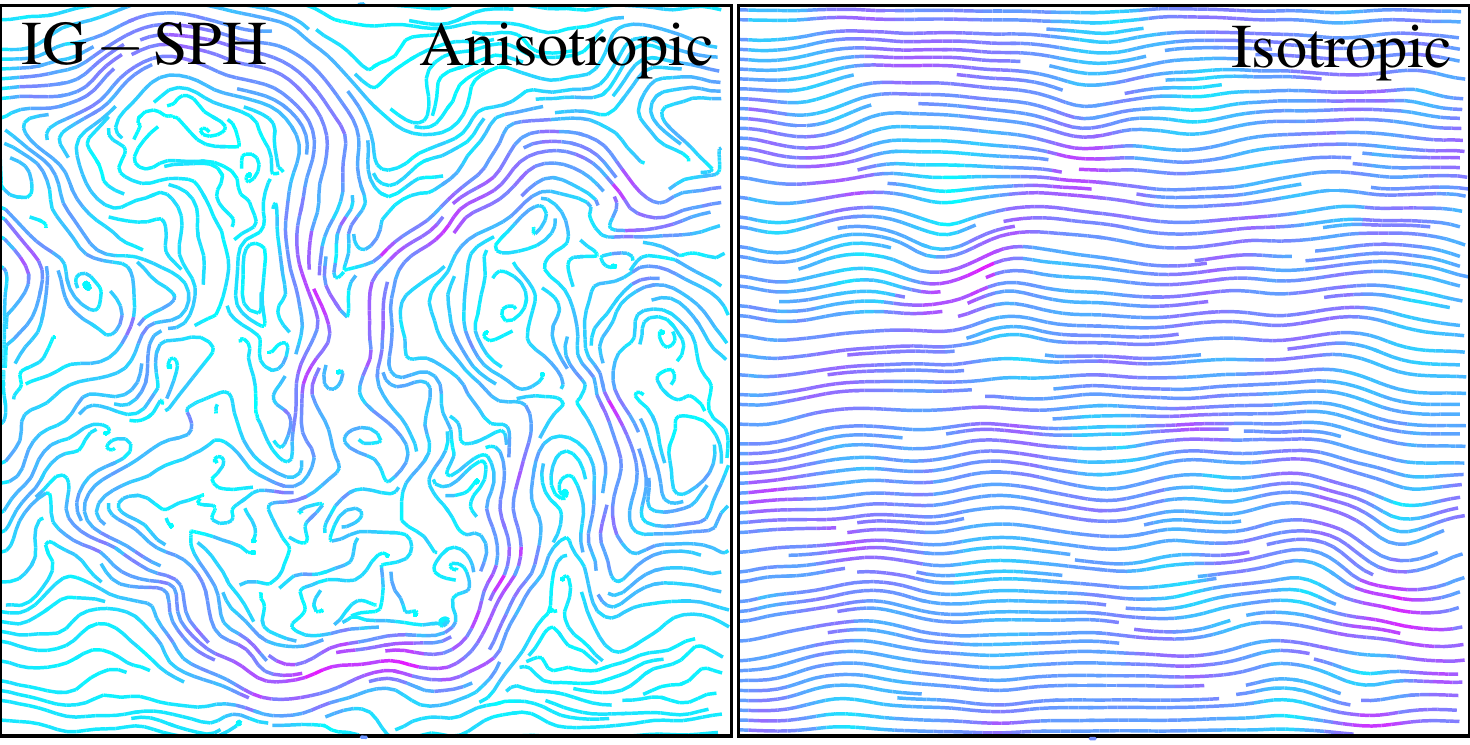} \\
  \includegraphics[width=0.9\columnwidth]{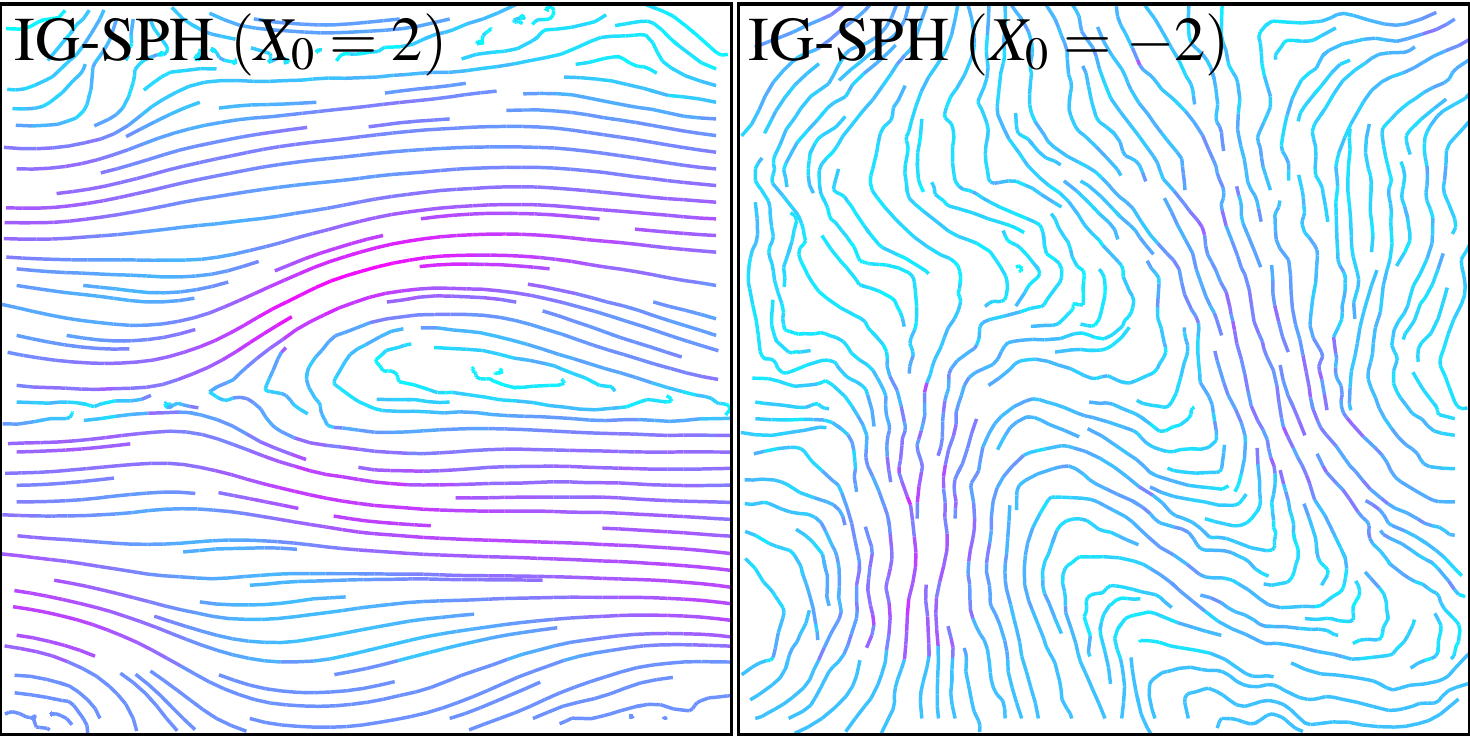} \\
   \end{tabular}
    \vspace{-0.25cm}
    \caption{{\em Top:} Diffusing ring test as Fig.~\ref{fig:ring}. 
    {\em Middle:} Low-resolution MTI test as Fig.~\ref{fig:mti.hbi}-\ref{fig:mti.time}, for both anisotropic and isotropic diffusion.
    {\em Bottom:} Hall MRI test as Fig.~\ref{fig:hall.mri.fieldlines}.
    IG-SPH performs comparably to our MFM method on the diffusing ring. In the dynamical problems (MTI and Hall MRI), 
    IG-SPH is able to capture some of the qualitative behavior, but known SPH challenges with highly sub-sonic turbulence produce the noise shown (and slower mode growth) and generate some instability even in the isotropic case (which should be stable). These may be fix-able, but are outside the scope of the diffusion operator itself.
        \vspacerpostplot 
    \label{fig:sph.multidimensional}}
\end{figure}

We can potentially improve on this, using our insight from the main text. For {\bf (1)}, replace the standard SPH estimator of $\langle \nabla \otimes {\bf q} \rangle_{a,\,{\rm SPH}}$ (Eq.~\ref{eqn:sph.first.deriv}) in Eq.~\ref{eqn:sph.grad.grad} with the second-order accurate moving least-squares estimator $(\nabla \otimes {\bf q})_{a}^{\rm matrix}$ from our MFM method (Eq.~\ref{eqn:gradient}). This is the same estimator used in so-called ``integral'' SPH formulations \citep{garciasenz:2012.integral.sph,rosswog:2014.sph.accuracy}. Unfortunately (as noted by those authors), this alone leads to a numerically unstable method. But we can remedy this by replacing the usual SPH fluxes with our flux-limited Riemann solution. To do this we modify Eq.~\ref{eqn:sph.grad.grad}, symmetrizing the density and smoothing length terms as they appear, giving: 
\begin{align}
\frac{d(V{\bf U})_{a}^{\rm SPH}}{dt} &= 
- \sum_{b}{\bf A}^{\rm SPH}_{ab} \cdot {\bf F}_{\rm diff}^{\rm SPH} \\ 
\label{eqn:sph.area} {\bf A}^{\rm SPH}_{ab} &\equiv \frac{m_{a}m_{b}}{\rho_{a}\rho_{b}}\,\left(\nabla_{a}W_{ab} + \nabla_{b}W_{ab}\right)\\ 
{\bf F}_{\rm diff}^{\rm SPH} &= -\frac{1}{2}\left( {\bf K}_{a}\cdot( \nabla \otimes {\bf q} )_{a}^{\rm matrix} + {\bf K}_{b}\cdot( \nabla \otimes {\bf q} )_{b}^{\rm matrix} \right)
\end{align}
and then replace ${\bf F}_{\rm diff}^{\rm SPH} \rightarrow {\bf F}_{\rm diff}^{\ast}$ solved exactly as in \S~\ref{sec:riemann}, identifying ${\bf F}_{{\rm diff},\,R} = {\bf K}_{b}\,( \nabla \otimes {\bf q} )_{b}^{\rm matrix}$ (and likewise $L$ and $a$). Note that this is now effectively a Godunov-type method, with the ``SPH effective face area'' ${\bf A}_{ab}^{\rm SPH}$. We therefore refer to this method as ``integral-Godunov'' (IG) SPH. Finally, note that this does not fully resolve issue {\bf (2)} above; doing so requires replacing ${\bf A}_{ab}^{\rm SPH}$ with a more consistent face estimator; of course, replacing this with our MFM faces gives {\em exactly} our MFM method in the main text (the only difference then would be how the other hydro equations are solved). 

Figs.~\ref{fig:sph.sheets}-\ref{fig:sph.multidimensional} show a series of tests of this modified IG-SPH method. First we simply verify the well-known conclusion that the HD-SPH form (Eq.~\ref{eqn:sph.eqn}) is unable to handle anisotropic diffusion. In contrast our IG-SPH method performs comparably to our MFM method (not surprising, given how similar they are) on simple diffusion sheet problems and e.g.\ the diffusing pulse and ring tests. We have also compared the ``gradients of gradients'' SPH method (Eq.~\ref{eqn:sph.grad.grad}); if the particle order is ideal this is comparable to our IG-SPH method, but if we seed initial noise in the field ${\bf q}$ or particle distribution, then it is significantly more noisy, and it fails to converge with resolution alone (keeping neighbor number fixed) on the diffusing ring test -- this is of course expected from our caveats {\bf (1)-(2)} above. Where IG-SPH shows some issues in Figs.~\ref{fig:sph.sheets}-\ref{fig:sph.multidimensional}: the Braginskii viscosity test, MTI, HBI, and Hall MRI tests, the problems appear to primarily owe to known difficulties with the SPH {\em hydrodynamics} operators (e.g.\ artificial viscosity triggering imperfectly, and low-order hydrodynamic errors generating noise in sub-sonic turbulence) -- there are various ways to improve these errors but since they are not in the diffusion operators themselves they are outside the scope of this paper.

\vspace{-0.5cm}
\section{Implicit Methods}
\label{sec:implicit}

In the main text we solve the diffusion equations explicitly. Qualitatively speaking, we update ${\bf U}$ according to 
\begin{align}
(V{\bf U})^{n+1} = (V{\bf U})^{n} + \Delta t\,\frac{d(V{\bf U})}{dt}^{n+1/2}
\end{align}
where $n$ refers to the timestep and $n+1/2$ refers to drifting all quantities to a half-timestep before calculating $d(V{\bf U})/dt$ (see \paperone\ for details). 

The explicit solver has the advantages of computational simplicity, extension to high-order gradient operators/reconstruction methods, allowing non-linear flux limiters, and trivially generalizing to hierarchical adaptive timesteps as we use in our code (akin to ``sub-cycling''). However, the disadvantage is that it requires a timestep limit of the form in Eq.~\ref{eqn:timesteps} (quadratic in the resolution). In some situations this can become extremely small; in these cases, implicit methods are popular. In the implicit case, we take 
\begin{align}
\label{eqn:implicit} (V{\bf U})^{n+1} = (V{\bf U})^{n} + \Delta t\,\frac{d(V{\bf U})}{dt}^{n+1}
\end{align}

\citet{petkova.springel.2009:otvet.gadget} show how to cast the isotropic SPH hybrid-derivatives diffusion equation in this form, and then re-cast it as a linear equation of the form ${\bf M}\cdot {\bf (VU)}^{n+1} = {\bf (VU)}^{n}$ for a vector ${\bf (VU)}$ of elements $(VU)_{a}$ (representing each resolution element $a$), and matrix ${\bf M}$. This can then be solved over all particles in a global timestep using a conjugate-gradient inversion. Unfortunately, for our MFM/MFV methods, deriving an implicit method is more challenging. If we simplify by using a first-order reconstruction of $\nabla \otimes {\bf q}$ (with corresponding limiters), and (for now) neglect the extra diffusion terms from the Riemann problem, and assume a scalar ${\bf q}$, then Eq.~\ref{eqn:faces} for $d(VU)/dt$ gives us
\begin{align}
(VU)_{a}^{n+1} =& (VU)_{a}^{n} \\
\nonumber &+ \frac{\Delta t}{2}\,\sum_{b}\,\left( {\bf K}_{a}\,\langle\nabla q\rangle^{n+1}_{a} + {\bf K}_{b}\,\langle\nabla q\rangle^{n+1}_{b} \right) \cdot \theta_{ab}\,{\bf A}_{ab}
\end{align}
where $\theta_{ab}$ is the limiter function based on comparing the implied flux here to the ``direct flux'' per Eq.~\ref{eqn:direct.limiter}. 
Combining this with Eq.~\ref{eqn:gradient} for the gradient estimators, re-arranging and simplifying, we obtain
\begin{align}
{\bf M}\cdot {\bf (VU)}^{n+1} =& {\bf (VU)}^{n} \\ 
\nonumber M_{ab} \equiv& \, \delta_{ab} + \frac{\Delta t}{2}\,\frac{\zeta_{b}}{V_{b}}{\Bigl[} 
{\bf K}_{b} \boldsymbol{\mu}_{b}^{S}\cdot \left( \tilde{\bf A}_{ab} + \tilde{\bf A}_{b}^{S}\,\delta_{ab}
\right) \\ 
& +\sum_{c} {\bf K}_{c}\boldsymbol{\mu}_{cb} \cdot \left( \tilde{\bf A}_{ac} - \tilde{\bf A}_{c}^{S}\,\delta_{ac} \right)
{\Bigr]} \\ 
\tilde{\bf A}_{ab} \equiv&\, \theta_{ab}\,{\bf A}_{ab} \ \ \ \ \ , \ \ \ \ \boldsymbol{\mu}_{ab} \equiv \beta_{a}{\bf W}^{-1}_{a}{\bf x}_{ba}\,\omega_{b}({\bf x}_{a}) \\ 
\tilde{\bf A}_{a}^{S} \equiv& \sum_{c}\tilde{\bf A}_{ab}\ \ \ \ , \ \ \ \ 
\boldsymbol{\mu}_{a}^{S} \equiv \sum_{c}\boldsymbol{\mu}_{ac}
\end{align}
where $\delta_{ab}$ is the Kronecker delta, $\omega$ and ${\bf W}^{-1}$ are defined in Eq.~\ref{eqn:gradient}, and $\beta$ is the slope-limiter for $\langle \nabla\otimes {\bf q} \rangle$ (see \paperone). Now, add the diffusive terms from the Riemann problem, ${\bf F} \rightarrow {\bf F} + {\bf F}_{\rm diss}$ where ${\bf F}_{\rm diss} = \alpha\,\lambda\,({\bf U}_{R}-{\bf U}_{L}) = \alpha_{ab}\,\lambda_{ab}\,[u_{a}-u_{b} + (\eta_{ab}/2)\,(\nabla u_{a} + \nabla u_{b})\cdot {\bf x}_{ba}]$ if we use a second-order reconstruction of  ${\bf U}_{R}$ and ${\bf U}_{L}$. Here $\lambda = |v_{R}-v_{L}|/2 + c_{\rm fast}$ and $\eta_{ab}$ is the appropriate limiter for the $U_{R}$, $U_{L}$ reconstruction, both as defined in the text. This gives us
\begin{align}
{\bf M}\rightarrow &\, {\bf M} + {\bf M}^{\rm diss} \\ 
\nonumber M^{\rm diss}_{ab} \equiv & - \frac{\Delta t}{2\,V_{b}}{\Bigl[}
2\left( \varphi_{ab} - \varphi_{b}^{S}\,\delta_{ab}\right) + 
\boldsymbol{\mu}_{b}^{S}\cdot \left( \boldsymbol{\Lambda}_{ab} + \boldsymbol{\Lambda}_{b}^{S}\,\delta_{ab} \right) \\ 
 & - \sum_{c} \boldsymbol{\mu}_{cb}\cdot \left(\boldsymbol{\Lambda}_{ac} + \boldsymbol{\Lambda}_{c}^{S} \delta_{ac} \right) {\Bigr]} \\ 
\varphi_{ab} \equiv&\alpha_{ab}\,\lambda_{ab}\,|\tilde{\bf A}|_{ab} \ \ \ \ , \ \ \ \ \boldsymbol{\Lambda}_{ab} \equiv \eta_{ab}\,\varphi_{ab}\,{\bf x}_{ba} \\ 
\varphi_{a}^{S} \equiv& \sum_{c}\varphi_{ac} \ \ \ \ \ \ \ \ \ \ \ \ , \ \ \ \ \boldsymbol{\Lambda}_{a}^{S} \equiv \sum_{c} \boldsymbol{\Lambda}_{ac}
\end{align}
The difference between MFM and IG-SPH simply amounts to the appropriate value for ${\bf A}_{ab}$.

If the limiter functions $\theta_{ab}$, $\eta_{ab}$, $\alpha_{ab}$, and $\beta_{a}$ were independent of $({\bf VU})$, then we could again perform a simple sparse-matrix inversion to update $({\bf VU})$. Obviously even in this limit, the matrix terms are quite complicated: note the extra summations over the index ``$c$'' that appear, which owe to the higher-order gradient estimator we adopt and cannot be eliminated without lowering the order of the gradient approximation. But the real problem is that the higher-order accuracy and stability of our MFM/MFV methods depends on the limiters being {\em non-linear} functions of {\em both} ${\bf U}_{a,\,b}$ and $\nabla \otimes {\bf q}_{a,\,b}$ (itself a sum over ${\bf U}_{a,\,b}$). Of course, non-linear global elliptical equations can be solved, but using an iterative root-finding method to solve for $({\bf VU})$, we would have to repeat the summation over ${\bf U}$ to determine $\nabla \otimes {\bf q}_{a,\,b}$ and re-compute the limiter functions between each iteration. Combined with a global timestep, this would make the implicit solver much more expensive than sub-cycling our explicit solver (defeating the purpose). 

In the limit where the problem is sufficiently smooth, well-resolved, and there is good particle order, $\theta,\,\eta,\,\beta\rightarrow1$ and $\alpha\rightarrow0$. Adopting these values and implementing the sparse matrix inversion, we are able to confirm our conclusions for the few test problems that satisfy these criteria. However, with these values fixed, the method is numerically unstable. We can stabilize the method for problems with unresolved gradients (still assuming good particle order) by taking $\theta,\,\beta,\,\alpha\rightarrow1$, $\eta\rightarrow0$, but this produces excessive numerical diffusion, especially at low resolution. Unfortunately, we see no obvious way to achieve the combination of accuracy and stability in the text (given the method studied here) without involving non-linear terms in $\nabla\otimes {\bf q}$, which are prohibitive for most implicit methods. However, it is possible that a semi-implicit method similar to the one in \citet{sharma.2011:fast.semi.implicit.anisotropic.diffusion} could be implemented, where a subset of the flux components which do not require strong limiters are solved implicitly while the others (where the limiters apply) are updated explicitly.

\vspace{-0.5cm}
\section{Super-Timestepping}
\label{sec:super.timestep}

Another method to allow larger timesteps in explicit methods for elliptic/parabolic equations is so-called ``super-timestepping'' \citep[see][and references therein]{alexiades.1996:super.timestepping,gurski.2010:supertimestep.rkc.implementation.nonsymmetric,meyer.2012:high.order.super.timestep.requires.grid,meyer.2014:higher.order.rkl.superstep.requires.grid}.

A super-timestep $\Delta t_{s}$ is composed of $N$ sub-steps $\delta t_{j}$
\begin{align}
\Delta t_{s} = \sum_{j=1}^{N}\,\delta t_{j}
\end{align}
The diffusion equations are calculated and updated as usual on each sub-step $\delta t_{j}$, however, these are essentially Runge-Kutta sub-steps in the sense that stability is not guaranteed at after any individual sub-step $\delta t_{j}$, but {\em only} on the super-step $\Delta t_{s}$. 

\citet{alexiades.1996:super.timestepping} show that the optimal timesteps which satisfy the necessary stability conditions on $\Delta t_{s}$ and simultaneously maximize $\Delta t_{s}$ for a given $N$ are given by
\begin{align}
\delta t_{j} = \Delta t_{\rm expl}\left[\left(1+\nu\right) - \left(1-\nu\right)\,\cos{\left(\frac{\pi\,(2j-1)}{2\,N} \right)} \right]^{-1}
\end{align}
where $\Delta t_{\rm expl}$ is our usual explicit timestep given by Eq.~\ref{eqn:timesteps} and $0 < \nu < \lambda_{\rm min}/\lambda_{\rm max}$ (where $\lambda_{\rm min}$ and $\lambda_{\rm max}$ are the smallest and largest eigenvalues of the diffusion equation). As $\nu\rightarrow 0$, the total time $\Delta t_{s}$ covered by the same number $N$ of substeps (and computations) increases, with the limit $\Delta t_{s} \rightarrow N^{2}\,\Delta t_{\rm expl}$ as $\nu\rightarrow 0$. Thus this method can provide a speedup of up to a factor $N$.

This is trivial to implement in explicit methods with a global timestep, and we have done so here. Although all problems shown in the text use the standard, non-optimized explicit timestep $\Delta t_{\rm expl}$, we have re-run all of them enforcing a global timestep and using the super-timestepping scheme with varying $N$ and $\nu$. Care is needed, since overly-aggressive timestepping can lead to a loss of accuracy even while maintaining stability (see discussion in \S~\ref{sec:timesteps.accelerated}). In practice, we find $N\sim 5-10$ and $\nu\sim 0.04$ give substantial speed-ups (factor $\sim 2.5-3$ on the problems here) without any measurable loss of accuracy (consistent with the studies in \citealt{alexiades.1996:super.timestepping,kim.2009:alexiades.super.timestep.ambipolar.diffusion} and \citealt{tsukamoto.2013:ohmic.dissipation.sph.super.timestepping}, who show good accuracy is maintained for $\nu \gtrsim 1/N^{1/2}$). 

Moreover, it is straightforward to generalize this to our individual timestepping scheme. In this scheme, particles have independent timesteps but are discretized into hierarchical powers-of-two timebins (see \paperone\ and \citealt{springel:gadget}); this ensures they remain synchronized. If we take each $\delta t_{j}$ and ``round down'' the nearest (smaller) power-of-two bin, it is straightforward to show that the stability condition in \citet{alexiades.1996:super.timestepping} is still satisfied (more easily, in fact). The tradeoff is that the timesteps are no longer quite ``optimal'' (not as large as possible). However, for careful choices of $N$ and $\nu$ (such as the values above), the difference in the super-step $\Delta t_{s}$ (the sum of the $\delta t_{j}$) is only $\sim 10\%$ below optimal.

Note that this is a first-order time integration scheme, which makes it particularly easy to implement. Still greater accuracy may be achieved with higher-order schemes following \citet{meyer.2012:high.order.super.timestep.requires.grid,meyer.2014:higher.order.rkl.superstep.requires.grid}, but we have not experimented with these.

\vspace{-0.5cm}
\section{Timestep Limits with Super-Timestepping or Implicit Solvers}
\label{sec:timesteps.accelerated}

With either implicit methods or super-timestepping, although formal stability is guaranteed for large time-steps, accuracy is not. For any test problem, we can simply vary the timestep until a desired accuracy is reached. However for more general problems, it is useful to have a more general timestep constraint. 

Regardless of the diffusion problem, MHD must obey the usual CFL condition $\Delta t < C_{\rm CFL}\,\Delta x / v_{\rm sig}$ (where $v_{\rm sig}$ is the standard signal velocity accounting for particle relative motions and their fastest wavespeeds). Additional physics (e.g.\ gravity, radiation, cooling) come with their own timestep constraints. Note that, for hyperbolic equations, we cannot circumvent these conditions using either of the acceleration schemes above; they are always enforced.

There is no rigorous equivalent condition for the diffusion equation with implicit/super-timestepped solvers, but as pointed out by many authors \citep[e.g.][]{kim.2009:alexiades.super.timestep.ambipolar.diffusion,gurski.2010:supertimestep.rkc.implementation.nonsymmetric,tsukamoto.2013:ohmic.dissipation.sph.super.timestepping}, motivated by hyperbolic problems, we can define an effective diffusive signal speed $v_{\rm diff} \sim \| {\bf K} \| \,\| \nabla \otimes {\bf q} \| /  \| {\bf U} \| $. Using this to define an analogous Courant condition on the super/implicit timestep: 
\begin{align}
\Delta t_{\rm implicit/super} < C_{\rm CFL}\,\Delta x\,\frac{ \| {\bf U} \| }{\| {\bf K} \| \,\| \nabla \otimes {\bf q} \|} \label{eqn:dt.implicit}
\end{align} 
we find reasonably good accuracy so long as $C_{\rm CFL} < 0.5$. For more details we refer to \citet{sharma.2010:cosmic.ray.streaming.timestepping} who rigorously demonstrate that such a criterion is the correct one for regimes where the diffusion equation behaves advectively (e.g.\ cosmic ray streaming, or our variable-diffusivity test problem). 

Note that when there are un-resolved gradients, this reduces (assuming ${\bf U}\sim {\bf q}$) to the usual explicit timestep-limiter, $\Delta t < \Delta x^{2} / \| {\bf K} \|$, as we might expect.\footnote{On some problems -- for example the diffusing sheet with perpendicular fields, implicit methods remain accurate even with much larger timesteps compared to Eq.~\ref{eqn:dt.implicit}. However, the only cases we find this is true are ones where the solution is steady-state or equilibrium (in which case arbitrarily long timesteps should not, in principle, be problematic). For all realistic problems we consider which feature actual dynamics, a criterion like Eq.~\ref{eqn:dt.implicit} is needed to maintain accuracy.} When the gradients are well-resolved, however, larger timesteps are allowed: for super-timestepping, this amounts to a speedup by a factor $\sim L_{\rm grad} / (N\,\Delta x)$ where $N$ is the number of sub-steps and $L_{\rm grad} = \| {\bf q} \| / \| \nabla \otimes {\bf q} \|$ is the gradient scale length. In practice, for most of the test problems in this paper, this translates to a maximum speedup of a factor of $\sim 3$ in super-timestepping before a noticeable loss of accuracy appears. Given the large overhead of the implicit methods, for the few cases where we can use them (e.g.\ ``standard'' SPH), we actually see relatively little speedup (factor $<2$). However, if we considered the same problems, in stages with {\em resolved} gradients, at much higher resolution, the difference between the simple explicit method used in the text and either implicit or super-timestepped methods should grow accordingly.

\end{appendix}

\end{document}